\documentclass[12pt]{iopart}
\usepackage{iopams}

\newcommand{\half}{\mbox{$\textstyle \frac{1}{2}$}}
\newcommand{\quat}{\mbox{$\textstyle \frac{1}{4}$}}
\newcommand{\octa}{\mbox{$\textstyle \frac{1}{8}$}}

\begin{document}
\title{Martingale Models for Quantum State Reduction}

\author{S.~L. Adler$^{*1}$, D.~C. Brody$^{\dagger2}$, T.~A.
Brun$^{*3}$ and L.~P. Hughston$^{\ddagger4}$}

\address{${}^{*}$Institute for Advanced Study, Einstein Drive,
Princeton, NJ 08540, USA}

\address{${}^{\dagger}$Blackett Laboratory, Imperial College,
London SW7 2BZ, UK}

\address{${}^{\ddagger}$Department of Mathematics, King's College
London, The Strand, London \\ WC2R 2LS, UK}

\begin{abstract}
Stochastic models for quantum state reduction give rise to
statistical laws that are in most respects in agreement with those
of quantum measurement theory. Here we examine the correspondence
of the two theories in detail, making a systematic use of the
methods of martingale theory. An analysis is carried out to
determine the magnitude of the fluctuations experienced by the
expectation of the observable during the course of the reduction
process and an upper bound is established for the ensemble average
of the greatest fluctuations incurred. We consider the general
projection postulate of L\"uders applicable in the case of a
possibly degenerate eigenvalue spectrum, and derive this result
rigorously from the underlying stochastic dynamics for state
reduction in the case of both a pure and a mixed initial state. We
also analyse the associated Lindblad equation for the evolution of
the density matrix, and obtain an exact time-dependent solution
for the state reduction that explicitly exhibits the transition
from a general initial density matrix to the L\"uders density
matrix. Finally, we apply Girsanov's theorem to derive a set of
simple formulae for the dynamics of the state in terms of a family
of geometric Brownian motions, thereby constructing an explicit
unravelling of the Lindblad equation.
\end{abstract}

\submitto{\JPA}

%
%

\section{Introduction}

According to nonrelativistic quantum mechanics, the evolution of
the state of an isolated quantum system is described by a
deterministic unitary transformation, governed by the
Schr\"odinger equation. The behaviour of the state of a quantum
system following the measurement of an observable is less well
understood, however, and has been the subject of much debate. In
quantum measurement theory it is usually assumed that when the
measurement of an observable with a discrete spectrum is carried
out on a system prepared in a prescribed initial state, then the
state reduces randomly to one of the eigenstates of the observable
being measured. This is the so-called projection postulate of von
Neumann \cite{neumann}, which was later generalised by L\"uders
\cite{luders} to handle the case of a measurement of an observable
with a degenerate spectrum. The L\"uders postulate has the virtue
of being unambiguously applicable whether or not the initial state
is pure, and whether or not the eigenvalue spectrum is
nondegenerate.

Stochastic extensions of the Schr\"odinger equation have been
increasingly attracting attention as plausible dynamical models
for state vector reduction in quantum mechanics
\cite{gisin,diosi,grw,percival,hughston,adler2,brody}. In such
models, the Schr\"odinger equation is generalised to take the form
of a special type of stochastic differential equation on Hilbert
space. For example, given a Hamiltonian ${\hat H}$ and a commuting
observable ${\hat F}$ with a discrete spectrum, there exists a
natural stochastic differential equation generalising the
Schr\"odinger equation with the property that starting from a
given initial pure state, the system evolves randomly in such a
way that asymptotically it reaches one of the eigenstates of
${\hat F}$ with the correct quantum probability. More generally,
given a compatible family of observables ${\hat F}_\alpha$
$(\alpha=1,2,\cdots,r)$, each of which commutes with ${\hat H}$,
then a similar result holds, with an asymptotic reduction to one
of the common eigenstates of the given family of observables. For
recent reviews outlining the development of this approach, with
extensive references, see \cite{pearle0} and \cite{ghirardi}.

The purpose of this paper is to analyse in some detail the
statistical laws associated with stochastic extensions of the
Schr\"odinger equation, and to show in particular how the
projection postulate in the general form due to L\"uders can be
derived as a consequence of the dynamics. The plan of the paper is
as follows. In sections 2 and 3 we review the von Neumann and
L\"uders versions of the projection postulate. In section 4 we
present a reasonably self-contained account of the basic
principles of stochastic state reduction, along with a brief
synopsis of the relevant mathematical tools of stochastic
analysis. For the purposes of illustration we consider primarily
the case of an energy-based reduction model, for which the
associated dynamics are given by the stochastic differential
equation (\ref{eq:3.1}), though most of the relevant mathematical
and physical ideas can be readily generalised to the class of
reduction processes noted above, based on a compatible family of
observables that commute with the Hamiltonian.

Throughout the discussion we emphasise the role of martingale
methods as an aid to the advancement of our understanding of
quantum phenomena. In particular we show that the expectation of
the Hamiltonian, which fluctuates during the course of the
reduction process, is a martingale, and its variance is a
supermartingale. The martingale property satisfied by the
expectation of the Hamiltonian can be viewed as a kind of weak
conservation law for the energy, generalising the Ehrenfest
relation. In section 5 we use the Doob-Kolmogorov maximal
inequalities to obtain a set of upper and lower bounds on the
fluctuations of the energy during the reduction process, and show
that the magnitude of a typical fluctuation is roughly of the
order of the initial energy uncertainty. We also demonstrate that
the process followed by the squared energy uncertainty is given by
the conditional variance of the terminal value of the energy.

In sections 6 and 7 we derive the projection postulate, and prove
that the collapse to the general L\"uders state occurs with the
appropriate probability, whether or not the initial state is pure
and whether or not the eigenvalue spectrum is degenerate. Then in
section 8 we study the implied evolution of the density matrix for
a given initial state, not necessarily pure, and derive an exact
time-dependent solution for the Lindblad equation associated with
the reduction. We conclude in section 9 by introducing a
change-of-measure technique to solve the stochastic differential
equation for the dynamics of the state vector, thereby
constructing an explicit unravelling of the Lindblad equation.

\section{Von Neumann Projection}

Let us consider a quantum system which, for mathematical
simplicity, we shall assume is characterised by a finite
dimensional Hilbert space ${\mathcal H}$ of dimension $N$. Suppose
that $F$ is an observable for which the corresponding Hermitian
operator acting on ${\mathcal H}$ is denoted ${\hat F}$. Our
interest here is in providing a clearer understanding of what
happens if the observable $F$ is measured when the system is in a
given pure state, corresponding to a ray through the origin in
${\mathcal H}$.

First we shall examine the more straightforward case where $F$ has
a nondegenerate spectrum, and the eigenvalues of ${\hat F}$ are
given by the numbers $f_n$ for $n=1,2,\cdots,N$, with the property
that $f_n\neq f_m$ for $n\neq m$. The eigenstate corresponding to
the eigenvalue $f_n$ will be denoted $|f_n\rangle$, so that ${\hat
F}|f_n\rangle=f_n|f_n\rangle$.

Let us write $|\psi_0\rangle$ for a representative state vector
for the given initial pure state. We shall assume that
$|\psi_0\rangle$ is normalised to unity, so $\langle\psi_0
|\psi_0\rangle =1$. We shall also assume that $\langle
f_n|f_m\rangle =\delta_{nm}$.

According to the {\sl projection postulate}, when a measurement of
$F$ is made, the state vector undergoes a transition
$|\psi_0\rangle \mapsto|f_n\rangle$ to one of the eigenstates of
${\hat F}$. This occurs for a specified value of $n$ with the
probability
\begin{eqnarray}
\pi_n=|\langle f_n|\psi_0\rangle|^2, \label{eq:2.1}
\end{eqnarray}
and the result of the measurement in that case is the eigenvalue
$f_n$. The associated transition $|\psi_0\rangle\mapsto
|f_n\rangle$ is called the state reduction or `collapse of the
wave function' arising from the measurement of $F$.

A more precise way of stating this is that when the observable $F$
is measured, the initial pure state $|\psi_0\rangle$ transforms to
a `mixture', i.e., a random state $|{\bf f}\rangle$ with the
property that $|{\bf f}\rangle$ is given by the eigenstate
$|f_n\rangle$ with the probability $\pi_n$. The fact that $|{\bf
f}\rangle$ is a mixture reflects our ignorance of what the result
of the measurement process will be. When the observable $F$ is
measured, we can be confident that the result is one of the values
$f_n$, and that the new state is the eigenstate represented by
$|f_n\rangle$, but we cannot say in advance which one it will be.
This is what is meant by saying that the result of the measurement
is random.

The density matrix ${\hat\rho}$ associated with the random state
$|{\bf f}\rangle$ and the probability distribution $\pi_n$ is
given by the expectation of the random projection operator $|{\bf
f}\rangle\langle{\bf f}|$ associated with $|{\bf f}\rangle$. In
other words, we have
\begin{eqnarray}
{\hat\rho} &=& {\mathbb E}\big[ |{\bf f}\rangle\langle{\bf f}|
\big] \nonumber \\ &=& \sum_{n=0}^{N} \pi_n |f_n\rangle\langle
f_n| . \label{eq:2.2}
\end{eqnarray}
Here ${\mathbb E}$ denotes expectation with respect to the
distribution $\pi_n$. The mixture $|{\bf f}\rangle$ typically
carries more information than the density matrix ${\hat\rho}$
alone, because if one is given ${\hat\rho}$, then in general there
are many different mixtures that correspond to it
\cite{hughston3}.

The significance of the density matrix ${\hat\rho}$ is that if $G$
is any other observable, not necessarily compatible with $F$, and
we measure $G$ after we measure $F$, then the expected value of
$G$ is
\begin{eqnarray}
{\mathbb E}\big[\langle{\bf f}|{\hat G}|{\bf f}\rangle\big] &=&
\sum_n \pi_n \langle f_n|{\hat G}|f_n \rangle \nonumber \\ &=&
{\rm Tr} \sum_n \pi_n |f_n\rangle\langle f_n|{\hat G} \\
\label{eq:2.3} &=& {\rm Tr}{\hat\rho}{\hat G}, \nonumber
\end{eqnarray}
where ${\rm Tr}$ denotes the trace operation.

An alternative way of facilitating the description of the density
matrix associated with the measurement outcome for an observable
$F$ is to introduce the projection operator ${\hat
P}_n=|f_n\rangle\langle f_n|$ associated with the eigenvalue
$f_n$. When $F$ is measured, the state $|\psi_0\rangle$ is
transformed to $\pi_n^{-1/2}{\hat P}_n |\psi_0\rangle$ if the
measurement outcome is known to be $f_n$, for which the
corresponding probability is $\pi_n$. An analogous transformation
holds for the corresponding density matrix ${\hat\rho}_0 =
|\psi_0\rangle\langle\psi_0|$ which transforms according to the
scheme
\begin{eqnarray}
{\hat\rho}_0\mapsto{\hat\rho}_\infty &=& \pi_n^{-1} {\hat P}_n
{\hat\rho}_0{\hat P}_n \nonumber \\ &=& |f_n\rangle\langle f_n| .
\label{eq:2.5}
\end{eqnarray}
Here we use the notation ${\hat\rho}_0$ to signify the density
matrix before the measurement, and ${\hat\rho}_\infty$ to signify
the density matrix after the measurement.

More generally, if $F$ is measured but the outcome is {\it not}
known, then ${\hat\rho}_0$ transforms according to the scheme
\begin{eqnarray}
{\hat\rho}_0\mapsto{\hat\rho}_\infty &=& \sum_{n=1}^{N} {\hat P}_n
{\hat\rho}_0 {\hat P}_n  \nonumber \\ &=& \sum_{n=1}^{N}
|f_n\rangle\langle f_n|\psi_0\rangle\langle\psi_0|f_n\rangle
\langle f_n| \label{eq:2.55} \\ &=& \sum_{n=1}^{N} \pi_n
|f_n\rangle\langle f_n| . \nonumber
\end{eqnarray}

The density matrix itself is often referred to as representing the
`state' of a quantum system. This is because it contains all the
information required to calculate ensemble expectations and
probabilities for the measurement outcomes of quantum observables,
conditional on the present state of knowledge of the observer.

In the examples above, the ensemble interpretation is as follows.
We prepare a large number of independent identical quantum systems
each in the state $|\psi_0\rangle$, then measure $F$. Now there
are two possibilities. In the first case, we only keep those
systems for which the result of the measurement of $F$ was the
value $f_n$. The new ensemble then consists of a large number of
independent systems each of which is in the pure state
$|f_n\rangle$. The corresponding density matrix is
$|f_n\rangle\langle f_n|$. In the second case, we keep all the
systems after $F$ has been measured. The resulting ensemble is
therefore a mixture, and for each value of $n$, a given system is
in the pure state $|f_n\rangle$ with probability $\pi_n$. The
corresponding density matrix is then given by (\ref{eq:2.55}).

\section{L\"uders' Postulate}

Let us now turn to the case of an observable with a degenerate
spectrum. In this case, we shall write $|f_{n,j}\rangle$ for an
orthogonal basis of distinct eigenstates of ${\hat F}$ sharing the
same eigenvalue $f_n$. Here $n=1,2,\cdots,D$, where $D$ is the
number of distinct energy levels, and  $j=1,2,\cdots,d_n$, where
$d_n$ is the dimensionality of the subspace ${\mathcal H}_n$ of
${\mathcal H}$ spanned by the eigenstates with eigenvalue $f_n$.
For convenience we normalise $|f_{n,j}\rangle$ such that
\begin{eqnarray}
\langle f_{n,j}|f_{m,k}\rangle = \delta_{nm}\delta_{jk} .
\label{eq:2.6}
\end{eqnarray}
Then the projection operator ${\hat P}_n$ onto the subspace
spanned by states for which $F=f_n$ is given by
\begin{eqnarray}
{\hat P}_n = \sum_{j=1}^{d_n}|f_{n,j}\rangle\langle f_{n,j}| .
\label{eq:2.7}
\end{eqnarray}
We note that ${\hat P}_n$ is independent of the specific choice of
basis made for the designated subspace, and that ${\hat P}_n{\hat
P}_m = {\hat P}_n \delta_{nm}$ and ${\hat P}_n |f_{m,k}\rangle =
\delta_{nm}|f_{m,k}\rangle$.

With these preliminaries in mind, suppose the observable $F$ is
measured when the system is in the pure state $|\psi_0\rangle$,
and the result of the measurement is one of the degenerate
eigenvalues $f_n$. In this case, it is perhaps less obvious {\it a
priori} what the correct probability is for the outcome, or indeed
what becomes of the state once the measurement result is known. A
more refined version of the projection postulate is required to
deal with this situation, due to L\"uders \cite{luders}, according
to which the measurement outcome probability is
\begin{eqnarray}
{\rm Prob}\left[ F=f_n\right] = \langle\psi_0|{\hat P}_n|\psi_0
\rangle , \label{eq:2.8}
\end{eqnarray}
and the associated state reduction is given by
\begin{eqnarray}
|\psi_0\rangle \mapsto \pi_n^{-1/2} {\hat P}_n |\psi_0\rangle,
\end{eqnarray}
where
\begin{eqnarray}
\pi_n &=& \langle{\psi}_0|{\hat P}_n |\psi_0\rangle \nonumber
\\ &=& \sum_{j=1}^{d_n} |\langle f_{n,j} |\psi_0\rangle|^2 .
\label{eq:2.9}
\end{eqnarray}
Thus, of all the possible eigenstates with eigenvalue $f_n$, a
single choice is made, given by the projection from the initial
state vector onto the relevant subspace. As in the nondegenerate
case, the measurement outcome can be described by a mixture, where
the random state $|{\bf f}\rangle$ is given by the normalised
L\"uders state $|f_n\rangle=\pi_n^{-1/2}{\hat P}_n|\psi_0\rangle$
with probability $\pi_n$ $(n=1,2,\cdots,D)$.

The validity of the L\"uders postulate can, in principle, be
tested by a succession of measurements of the energy of a system,
followed by the measurement of another observable incompatible
with the energy. Consider for example the system consisting of a
pair of noninteracting spin-$\frac{1}{2}$ particles in an external
magnetic field aligned along the $z$-axis, for which the energy
eigenstates are the spin-0 singlet and spin-1 triplet. The
corresponding eigenvalues are degenerate and are given by
$E_1=-1$, $E_2=E_3=0$, and $E_4=1$, in suitable units. We choose
the initial state such that the L\"uders state associated with the
degenerate energy eigenvalue is given by the spin-0 singlet. Then,
for an ensemble of identically prepared systems, we measure the
energy, and discard those systems for which the outcome is given
by one of the eigenvalues $\pm1$. The remaining systems, according
to L\"uders, are in the spin-0 singlet state, whereas according to
von Neumann \cite{neumann}, these states would in general be in
superpositions of the singlet and $S_z=0$ triplet states, the
precise details of which depend on the nature of the measurement
apparatus. In the present set-up, one can measure the total spin
operator to determine the outcome of the initial energy
measurement. This is because the result of the total spin gives
the eigenvalue $0$ if and only if the system is in the singlet
state.

An important feature of the L\"uders postulate is the inherent
`instability' of the reduction process implied for certain types
of measurements. That is, in the case of an observable with a
degenerate eigenvalue $f_n$, the projection is onto a single state
$\pi_n^{-1/2}{\hat P}_n|\psi_0\rangle$; whereas if the observable
is perturbed even slightly, breaking the degeneracy and producing,
say, two distinct but close eigenvalues $f_n$ and $f_{n'}$, then
the reduction process bifurcates, leading to one or the other of
two orthogonal, and hence maximally separated, eigenstates
$|f_n\rangle$ and $|f_{n'}\rangle$.

Thus we are led to consider that if the phenomenon of state
reduction itself arises as a consequence of a {\sl dynamical
process}, then this process must have sufficiently special
properties to ensure that under a smooth deformation of the
parameters characterising the observable being measured, the
resulting dynamics exhibits the required discontinuous behaviour
and produces the corresponding bifurcation in the postulated
transition probabilities. In what follows, we shall demonstrate
that the standard stochastic models for quantum state reduction
exhibit the required special properties, including the relevant
bifurcation phenomena. In short, we can {\sl derive} the
projection postulate from a dynamical model, including the
specifics of the L\"uders postulate and its consequences in the
case of a degenerate spectrum and a mixed initial state.

\section{State Vector Reduction}

In this section we consider in some detail the case of the
standard energy-based stochastic extension of the Schr\"odinger
equation, for which, if we set $\hbar=1$, the dynamics are given
by the following stochastic differential equation on ${\mathcal
H}$:
\begin{eqnarray}
{\rm d}|\psi_t\rangle &=& -{\rm i}{\hat H} |\psi_t\rangle {\rm d}t
- \octa \sigma^2({\hat H}-H_t)^2 |\psi_t\rangle {\rm d}t + \half
\sigma ({\hat H}-H_t)|\psi_t\rangle {\rm d}W_t . \label{eq:3.1}
\end{eqnarray}
The properties of dynamical processes of this type have been
investigated by a number of authors
\cite{diosi,percival,hughston,adler2,brody}. In particular, as we
shall demonstrate with various examples, many of the familiar
probabilistic features of standard quantum mechanics, including
the Born rules, can be {\sl deduced} from (\ref{eq:3.1}), or
suitable generalisations thereof. Here, $|\psi_t\rangle$ denotes
the random state vector at time $t$, for which an initial state
$|\psi_0\rangle$ is prescribed. For the moment we consider the
case when $|\psi_0\rangle$ is known, and later we turn to the case
where $|\psi_0\rangle$ is random. For the expectation of the
Hamiltonian operator ${\hat H}$ in the state $|\psi_t\rangle$ we
write
\begin{eqnarray}
H_t = \frac{\langle{\psi}_t|{\hat H}|\psi_t\rangle}
{\langle{\psi}_t|\psi_t\rangle} , \label{eq:3.2}
\end{eqnarray}
from which it follows that $H_t$ itself can be interpreted as a
random process, which we shall call the {\sl energy process}.
Strictly speaking, an expression such as $({\hat H}-H_t)^2
|\psi_t\rangle$ in (\ref{eq:3.1}) should be written $({\hat
H}-H_t{\hat 1})^2 |\psi_t\rangle$, where ${\hat 1}$ is the
identity operator, but there will be no ambiguity if we use the
more compact notation.

The energy-based stochastic extension of the Schr\"odinger
equation (\ref{eq:3.1}) is of great interest because it represents
perhaps the simplest plausible model for the collapse of the wave
function, and as such exhibits many remarkable features, both
physically and mathematically. It will be useful if we begin our
analysis with a brief overview of the probabilistic framework
implicit in the characterisation of (\ref{eq:3.1}).

The stochastic differential equation for the process
$|\psi_t\rangle$ is to be understood as defined on a fixed
probability space $(\Omega,{\mathcal F},{\mathbb P})$ equipped
with a filtration ${\mathcal F}_t$, with respect to which $W_t$ is
a standard Wiener process (Brownian motion). Here $\Omega$ is the
sample space, ${\mathcal F}$ is a $\sigma$-field on $\Omega$, and
${\mathbb P}$ is a probability measure on ${\mathcal F}$.

The filtration ${\mathcal F}_t$ represents the information
available at time $t$, where $t\in[0,\infty)$. More precisely, a
filtration of ${\mathcal F}$ is a collection ${\mathcal F}_t$
$(0\leq t<\infty)$ of $\sigma$-subfields of ${\mathcal F}$ with
the property that $s\leq t$ implies ${\mathcal F}_s\subset
{\mathcal F}_t$. A function $X: \Omega\mapsto {\mathbb R}$ is said
to be {\sl measurable} with respect to ${\mathcal F}$ if for each
$x\in{\mathbb R}$ the set consisting of all $\omega\in\Omega$
satisfying $X(\omega)\leq x$ is an element of ${\mathcal F}$. This
assures that ${\rm Prob}[X\leq x]$ exists with respect to the
given measure ${\mathbb P}$ on ${\mathcal F}$, and we say that $X$
is a random variable on the probability space $(\Omega,{\mathcal
F},{\mathbb P})$. Then by a random process we mean a parameterised
family of random variables $X_t$ $(0\leq t<\infty)$ on $(\Omega,
{\mathcal F},{\mathbb P})$. With a slight abuse of notation we let
$X_t$ stand both for the entire process $X_t$ $(0\leq t<\infty)$,
as well as the random variable $X_t$ for some given value of $t$;
usually it will be evident from context which meaning is intended.
Likewise ${\mathcal F}_t$ may denote the entire filtration
${\mathcal F}_t$ $(0\leq t<\infty)$, or the $\sigma$-subfield of
${\mathcal F}$ corresponding to the information set at time $t$.
If a random process $X_t$ is such that for each value of $t$ the
corresponding random variable $X_t$ is ${\mathcal
F}_t$-measurable, then we say that the process $X_t$ is {\sl
adapted} to the filtration ${\mathcal F}_t$. The idea of
`adaptedness' is important because it is though this device that a
notion of causality is introduced for the class of process we
consider.

It should be emphasised that the probabilistic concepts outlined
here and in what follows are introduced not merely for the sake of
mathematical clarity (although this is in itself a desirable
feature), but also because it makes possible the use of various
powerful analytical tools, examples of which we shall discuss
shortly.

The concept of conditional expectation plays a particularly
important role in the theory of quantum state reduction, and hence
it will be helpful if we elaborate slightly on the idea here. For
any random variable $X$ on $(\Omega,{\mathcal F},{\mathbb P})$ we
define its expectation
\begin{eqnarray}
{\mathbb E}[X] = \int_{\Omega} X(\omega){\rm d}{\mathbb P}(\omega)
\end{eqnarray}
by use of the Lebesgue integral. One can think of ${\mathbb E}[X]$
as the ensemble average of $X$. Then if ${\mathcal E}$ is a
$\sigma$-subfield of ${\mathcal F}$, the random variable $Y$ is
said to be (a version of) the conditional expectation of $X$ with
respect to ${\mathcal E}$ if $Y$ is ${\mathcal E}$-measurable and
if ${\mathbb E}[X1_{A}]={\mathbb E}[Y1_A]$ for all sets $A\in
{\mathcal E}$. In that case we write $Y={\mathbb E}[X|{\mathcal
E}]$. Here $1_A$ denotes the indicator function for the set $A$,
so $1_A(\omega)=1$ for $\omega\in A$ and $1_A(\omega)=0$ for
$\omega\notin A$. By convention we write ${\mathbb E}_t[X] =
{\mathbb E}[X|{\mathcal F}_t]$ for the conditional expectation of
$X$ given information up to time $t$. One can think of ${\mathbb
E}_t[X]$ as the ensemble average of $X$ conditional on the history
of events up to time $t$ being specified.

A useful result that follows on from these definitions is the
so-called `tower property' of conditional expectation, which says
that if ${\mathcal D}$ is a $\sigma$-subfield of ${\mathcal E}$
then ${\mathbb E}\left[ {\mathbb E}[X|{\mathcal E}]|{\mathcal D}
\right] = {\mathbb E}[X|{\mathcal D}]$. If we set ${\mathcal D}=
(\emptyset,\Omega)$, the smallest $\sigma$-subfield of ${\mathcal
F}$, then ${\mathbb E}[X|{\mathcal D}]= {\mathbb E}[X]$. This is
because the only ${\mathcal D}$-measurable random variables in
that case are the constants: if $X(\omega)$ is constant on
$\Omega$, then for any given $x$ we have $X(\omega)\leq x$ either
for all $\omega\in\Omega$ or for no $\omega\in\Omega$; conversely,
if $X(\omega)$ is not constant, then we can find a value of $x$
and two points $\omega_1,\omega_2\in\Omega$ such that
$X(\omega_1)\leq x$ and $X(\omega_2)>x$, which shows that
$X(\omega)$ is not ${\mathcal D}$-measurable---that is to say, the
set $\{\omega:\ X(\omega)\leq x\}$ is not an element of ${\mathcal
D}$. It follows then from the tower property that ${\mathbb E}
\left[ {\mathbb E}[X|{\mathcal E}]\right] = {\mathbb E}[X]$, the
so-called {\sl law of total probability}. In the case of a
filtration ${\mathcal F}_t$ $(0\leq t<\infty)$, if we take
${\mathcal E}={\mathcal F}_t$ and ${\mathcal D}={\mathcal F}_s$,
then the tower property reads ${\mathbb E}_s\left[ {\mathbb E}_t
[X] \right] = {\mathbb E}_s[X]$ for $s\leq t$; whereas the law of
total probability implies that ${\mathbb E}\left[ {\mathbb E}_t
[X] \right] = {\mathbb E}[X]$ for all $t\geq0$.

Now suppose $X_t$ $(0\leq t<\infty)$ is an adapted process on a
probability space $(\Omega,{\mathcal F},{\mathbb P})$ with
filtration ${\mathcal F}_t$ $(0\leq t<\infty)$. Then we say $X_t$
is a {\sl martingale} if the following two conditions hold:
${\mathbb E}[|X_t|]<\infty$ for all $0\leq t<\infty$, and
${\mathbb E}_s[X_t]=X_s$ for all $0\leq s\leq t<\infty$. If
instead of the latter condition $X_t$ satisfies ${\mathbb
E}_s[X_t]\geq X_s$ then we say $X_t$ is a {\sl submartingale}, and
if ${\mathbb E}_s[X_t]\leq X_s$ we say $X_t$ is a {\sl
supermartingale}.

Returning now to our investigation of the process (\ref{eq:3.1}),
we note that the parameter $\sigma$ governs the rate at which the
state vector reduction proceeds for a given level of initial
uncertainty in the energy. The units of $\sigma$ are
\begin{eqnarray}
\sigma\sim[{\rm energy}]^{-1}[{\rm time}]^{-1/2}. \label{eq:3.3}
\end{eqnarray}
The characteristic time-scale $\tau_R$ associated with the
collapse of the wave function is
\begin{eqnarray}
\tau_R = \frac{1}{\sigma^2 V_0} ,\label{eq:3.4}
\end{eqnarray}
where $V_0$ is the square of the initial energy uncertainty
$\Delta H$.

In what follows we shall make no specific assumptions about the
value of $\sigma$. Nevertheless, we note, as is discussed in
\cite{hughston,adler2}, that if $\sigma\sim M_p^{-1/2}$ in
microscopic units with $\hbar=c=1$, where $M_p$ is the Planck
mass, then the `large numbers' cancel out and we are left with a
typical reduction time-scale of
\begin{eqnarray}
\tau_R \sim \left( \frac{2.8{\rm MeV}}{\Delta H}\right)^2 {\rm s}.
\end{eqnarray}
This expression is interesting as a candidate for $\tau_R$
inasmuch as it relates energy spreads typical of atomic and
nuclear phenomena to time-scales that are accessible in the
laboratory.

The factor of $\frac{1}{2}$ appearing in front of $\sigma$ in
(\ref{eq:3.1}) is for convenience, and ensures consistency with
the notation of \cite{hughston,adler2,brody}.

It follows from the Ito rules $({\rm d}t)^2=0$, ${\rm d}t{\rm
d}W_t=0$, and $({\rm d}W_t)^2={\rm d}t$, as well as the special
form of the nonlinear terms appearing in (\ref{eq:3.1}), that the
norm of the state $|\psi_t\rangle$ is preserved under the
evolution (\ref{eq:3.1}). This can be seen as follows. The Ito
product rule states that if $X_t$ and $Y_t$ are Ito processes then
${\rm d}(X_tY_t) = Y_t {\rm d}X_t + X_t {\rm d}Y_t + {\rm d} X_t
{\rm d}Y_t$. As a consequence, we have
\begin{eqnarray}
{\rm d}(\langle{\psi}_t|\psi_t\rangle) = ({\rm d}\langle
{\psi}_t|)|\psi_t\rangle + \langle{\psi}_t|({\rm d}
|\psi_t\rangle) + ({\rm d}\langle{\psi}_t|) ({\rm d}
|\psi_t\rangle) \label{eq:3.42} .
\end{eqnarray}
Now the Hermitian conjugate of (\ref{eq:3.1}) is
\begin{eqnarray}
{\rm d}\langle{\psi}_t| = {\rm i} \langle{\psi}_t| {\hat H} {\rm
d}t - \octa\sigma^2 \langle{\psi}_t|({\hat H}-H_t)^2 {\rm d}t +
\half\sigma \langle{\psi}_t|({\hat H}-H_t) {\rm d} W_t
\label{eq:3.43}
\end{eqnarray}
Therefore, by use of the Ito rules, we obtain
\begin{eqnarray}
({\rm d}\langle{\psi}_t|)|\psi_t\rangle = \left( {\rm i} H_t -
\octa\sigma^2 V_t\right) \langle\psi_t|\psi_t\rangle{\rm d}t ,
\end{eqnarray}
and its conjugate,
\begin{eqnarray}
\langle{\psi}_t|({\rm d}|\psi_t\rangle) = \left( -{\rm i} H_t -
\octa\sigma^2 V_t\right) \langle\psi_t|\psi_t\rangle{\rm d}t ,
\end{eqnarray}
together with
\begin{eqnarray}
({\rm d}\langle{\psi}_t|)({\rm d}|\psi_t\rangle) = \quat \sigma^2
V_t \langle\psi_t|\psi_t\rangle{\rm d}t ,
\end{eqnarray}
where $V_t$ is given by formula (\ref{eq:3.6}) below. It follows
then from (\ref{eq:3.42}) that ${\rm d} (\langle {\psi}_t
|\psi_t\rangle)=0$, as desired. This result is useful in
calculations because we can assume the initial norm to be unity,
without loss of generality, and thus $\langle{\psi}_t
|\psi_t\rangle=1$ for all $t$.

An analogous calculation shows that the energy  process $H_t$
defined in (\ref{eq:3.2}) satisfies
\begin{eqnarray}
{\rm d}H_t = \sigma V_t {\rm d}W_t , \label{eq:3.5}
\end{eqnarray}
where $V_t$ is the process for the variance (squared uncertainty)
of ${\hat H}$ in the state $|\psi_t\rangle$, given by
\begin{eqnarray}
V_t = \frac{\langle{\psi}_t|({\hat H}-H_t)^2
|\psi_t\rangle}{\langle{\psi}_t|\psi_t\rangle} . \label{eq:3.6}
\end{eqnarray}
The variance process for the Hamiltonian has the property that
$V_t=0$ at time $t$ if and only if $|\psi_t\rangle$ is an energy
eigenstate at that time. As a consequence of (\ref{eq:3.5}), we
can write
\begin{eqnarray}
H_t = H_0 + \sigma \int_0^t V_u {\rm d}W_u , \label{eq:3.7}
\end{eqnarray}
where $H_0$ is the initial expectation value for the energy. Now
it is a general property of the stochastic integral that for any
${\mathcal F}_t$-adapted integrand $A_t$ satisfying ${\mathbb E}
\left[ \int_0^t A_u^2 {\rm d}u\right] <\infty$ we have
\begin{eqnarray}
{\mathbb E}_s\left[ \int_0^t A_u {\rm d}W_u \right] = \int_0^s A_u
{\rm d}W_u \quad\quad (s\leq t) \label{eq:3.75}.
\end{eqnarray}
The variance process $V_t$ is bounded by $\frac{1}{4}(E_+-E_-)^2$
where $E_+$ and $E_-$ are the largest and the smallest energy
eigenvalues, respectively, which implies that $\int_0^t V_u^2 {\rm
d}u<\infty$. Furthermore, we note that $|H_t|$ is bounded by
$\max(|E_+|, |E_-|)$. It follows that $H_t$ is a martingale:
\begin{eqnarray}
{\mathbb E}_s [H_t] = H_s, \quad\quad (s\leq t) . \label{eq:3.8}
\end{eqnarray}
The martingale condition is the stochastic analogue of a
conservation law, and thus (\ref{eq:3.8}) can be interpreted as a
{\it weak conservation law for the energy}. We recall that, for
ordinary quantum-mechanical evolution in the case of a
time-independent Hamiltonian, the Schr\"odinger equation
$\partial_t|\psi_t\rangle=-{\rm i}{\hat H}|\psi_t\rangle$ ensures
that the expectation of the Hamiltonian $H_t=\langle{\psi}_t|{\hat
H} |\psi_t\rangle /\langle{\psi}_t|\psi_t\rangle$ is conserved
along the Schr\"odinger trajectories. In the case of the
stochastic extension of the Schr\"odinger equation we have instead
the martingale relation (\ref{eq:3.8}) which ensures that the
ensemble average of the energy is conserved.

Because (\ref{eq:3.8}) plays a pivotal role in understanding the
nature of the reduction process, we shall sometimes refer to the
system of stochastic dynamics described by (\ref{eq:3.1}) as a
{\sl martingale model} for quantum state reduction.

We note, more generally, that if the operator ${\hat G}=g({\hat
H})$ is given by a function of ${\hat H}$, then the process
\begin{eqnarray}
G_t = \frac{\langle{\psi}_t|{\hat G}|\psi_t\rangle}
{\langle{\psi}_t|\psi_t\rangle} \label{eq:3.85}
\end{eqnarray}
is also weakly conserved, i.e., ${\mathbb E}_s[G_t]=G_s$ for
$s\leq t$. Thus, for example, if $g(x)=x^n$ and we introduce the
notation
\begin{eqnarray}
H^{(n)}_t = \frac{\langle{\psi}_t|{\hat H}^n|\psi_t\rangle}
{\langle{\psi}_t|\psi_t\rangle} \label{eq:3.851}
\end{eqnarray}
for the $n$-th moment of the energy, then
\begin{eqnarray}
{\rm d}H^{(n)}_t = \sigma\left( H^{(n+1)}_t-H_tH^{(n)}_t\right)
{\rm d}W_t , \label{eq:3.852}
\end{eqnarray}
where $H_t=H_t^{(1)}$.

With these formulae in mind, let us consider now the dynamics of
the variance process $V_t$. Writing $V_t=H^{(2)}_t-(H_t)^2$ it
follows, according to Ito's lemma, that
\begin{eqnarray}
{\rm d}V_t = {\rm d}H^{(2)}_t-2H_t{\rm d}H_t - ({\rm d}H_t)^2 .
\label{eq:3.853}
\end{eqnarray}
By use of the Ito rules together with (\ref{eq:3.852}) we then
deduce that
\begin{eqnarray}
{\rm d}V_t = -\sigma^2 V_t^2 {\rm d}t + \sigma \beta_t {\rm d}W_t
, \label{eq:3.9}
\end{eqnarray}
where
\begin{eqnarray}
\beta_t = \frac{\langle{\psi}_t|({\hat H}-H_t)^3
|\psi_t\rangle}{\langle{\psi}_t|\psi_t\rangle}
\end{eqnarray}
is the {\sl skewness} of the energy distribution at time $t$,
i.e., the third central moment of the Hamiltonian. More
specifically, we have $\beta_t = H^{(3)}_t
-3H_tH^{(2)}_t+2(H_t)^3$.

Integrating equation (\ref{eq:3.9}) for the dynamics of the
variance we obtain
\begin{eqnarray}
V_t = V_0 -\sigma^2 \int_0^t V_u^2 {\rm d}u + \sigma \int_0^t
\beta_u {\rm d}W_u , \label{eq:3.11}
\end{eqnarray}
from which it follows at once, by use of (\ref{eq:3.75}), that
\begin{eqnarray}
{\mathbb E}_s [V_t] = V_s - \sigma^2 {\mathbb E}_s \left[ \int_s^t
V_u^2 {\rm d}u \right], \label{eq:3.115}
\end{eqnarray}
and thus
\begin{eqnarray}
{\mathbb E}_s [V_t] \leq V_s \label{eq:3.12}
\end{eqnarray}
for $s\leq t$, which shows that $V_t$ is a {\sl supermartingale},
i.e., a process that on average decreases.

In particular, if we write ${\bar V}_t={\mathbb E}[V_t]$ for the
ensemble average of $V_t$, then it follows as a special case of
(\ref{eq:3.115}) that
\begin{eqnarray}
{\bar V}_t = V_0 - \sigma^2 {\mathbb E}\left[ \int_0^t V_u^2 {\rm
d}u \right] . \label{eq:3.123}
\end{eqnarray}
Differentiating this expression with respect to $t$ we obtain
\begin{eqnarray}
\frac{{\rm d}{\bar V}_t}{{\rm d}t} = -\sigma^2 {\bar V}_t^2
(1+\eta_t) , \label{eq:3.13}
\end{eqnarray}
where the process $\eta_t$ defined by $\eta_t = {\mathbb E}
[(V_t-{\bar V}_t)^2]/{\bar V}_t^2$ is nonnegative. Therefore, by
integration of (\ref{eq:3.13}), we obtain
\begin{eqnarray}
{\bar V}_t = \frac{V_0}{1+\sigma^2V_0(t+\xi_t)}, \label{eq:3.15}
\end{eqnarray}
where $\xi_t=\int_0^t \eta_s {\rm d}s$. As a consequence we deduce
that
\begin{eqnarray}
{\bar V}_t \leq \frac{V_0}{1+\sigma^2V_0t} , \label{eq:3.16}
\end{eqnarray}
which shows that
\begin{eqnarray}
\lim_{t\rightarrow\infty}{\bar V}_t=0.
\end{eqnarray}
Alternatively, if we introduce the `localisation' process
$\Lambda_t={\bar V}_t^{-1}$ then it follows from (\ref{eq:3.13})
that $\partial_t\Lambda_t \geq \sigma^2$, which shows that
$\Lambda_t$ increases without bound \cite{percival3,schack}. In
(\ref{eq:3.16}) we see an example of the role of
$\tau_R=(\sigma^2V_0)^{-1}$ as the characteristic time-scale of
the reduction process. Since $V_t$ is nonnegative, it follows that
\begin{eqnarray}
\lim_{t\rightarrow\infty}V_t=0 \label{eq:3.161}
\end{eqnarray}
almost surely. The dynamical process (\ref{eq:3.1}) therefore
induces a collapse of the wave function, for any choice of the
initial state $|\psi_0\rangle$, to an eigenstate of the
Hamiltonian.

\section{Fluctuation Analysis}

The martingale property (\ref{eq:3.8}) satisfied by the energy
process $H_t$ implies in the limit $t\rightarrow\infty$, that
${\mathbb E} [H_\infty]= H_0$. However, the terminal value
$H_\infty$ of the energy process, the existence of which we shall
establish shortly, is necessarily one of the energy eigenvalues,
from which it follows that
\begin{eqnarray}
H_0 = \sum_n \pi_n E_n , \label{eq:3.18}
\end{eqnarray}
where $\pi_n$ is the probability of reaching the eigenstate
$|n\rangle$ starting from the given initial state. Therefore, the
ensemble average of the measured value of the energy equals the
expectation value of the energy in the initial state, as it
should.

The importance of this conclusion is that whereas in quantum
measurement theory it is essentially an {\it assumption} that the
`expectation value' of an observable in a given state is the
ensemble average for the result of a measurement of the
observable, in a martingale model one can {\it prove} that the
asymptotic ensemble average agrees with the expectation value,
hence justifying the conventional interpretation of this quantity.
In particular, using (\ref{eq:3.8}) we can write
\begin{eqnarray}
H_t = {\mathbb E}_t[H_\infty]
\end{eqnarray}
which shows that the quantum expectation value $H_t$ of the
observable $H$ at time $t$ is always the best `forecast', based on
information currently available, for the outcome of a measurement
of $H$.

A similar result holds for the dispersion of the measured values
of the energy. This can be established by use of the Ito isometry.
If $A_t$ and $B_t$ are ${\mathcal F}_t$-adapted real processes
that are square-integrable in the sense that ${\mathbb E}
[\int_0^t A_s^2{\rm d}s]<\infty$ and ${\mathbb
E}[\int_0^tB_s^2{\rm d}s]<\infty$, then the Ito isometry states
that
\begin{eqnarray}
{\mathbb E}\left[\left(\int_0^tA_s {\rm d}W_s \right)\left(
\int_0^tB_s {\rm d}W_s\right)\right] = {\mathbb E} \left[
\int_0^tA_sB_s{\rm d}s\right] . \label{eq:3.191}
\end{eqnarray}
It follows therefore from (\ref{eq:3.7}) that
\begin{eqnarray}
{\mathbb E}\left[ (H_t-H_0)^2 \right] &=& \sigma^2 {\mathbb E}
\left[ \left( \int_0^t V_s {\rm d}W_s \right)^2 \right] \nonumber
\\ &=& \sigma^2 {\mathbb E}\left[ \int_0^t V_s^2 {\rm d}s \right]
\label{eq:3.19}
\end{eqnarray}
by virtue of the Ito isometry. By use of expression
(\ref{eq:3.11}) for $V_t$ we then deduce that
\begin{eqnarray}
{\mathbb E}\left[ (H_t-H_0)^2 \right] = V_0 - {\mathbb E} [V_t] .
\label{eq:3.20}
\end{eqnarray}
Taking the limit $t\rightarrow\infty$ and using the fact that
$\lim_{t\rightarrow\infty}{\bar V}_t=0$, we get
\begin{eqnarray}
{\rm Var}\left[ H_\infty\right] &=& {\mathbb E}\left[ (H_\infty -
{\mathbb E}[H_\infty])^2 \right] \nonumber \\ &=& V_0 ,
\label{eq:3.21}
\end{eqnarray}
which demonstrates that {\it the variance of the measured energy
is in agreement with the squared energy uncertainty in the initial
state}.

During the course of the reduction process, the energy $H_t$ of
the system can, in principle, deviate far from its initial value
$H_0$, subject to the condition that it stays in the range
$H_t\in[E_-,E_+]$, where $E_-$ and $E_+$ are the lowest and
highest energy levels. Nevertheless, we can show that on average
$H_t$ will not deviate too much from $H_0$: an upper bound can be
set on the maximum fluctuation experienced by the energy, on
average, as the reduction proceeds. This bound is given by
$2\Delta H$, twice the initial energy uncertainty.

The proof of this result makes use of the Doob-Kolmogorov maximal
inequalities (see, e.g., \cite{ikeda}, theorem 6.10, or
\cite{yor}, theorem 1.7, p. 54). These inequalities state that if
$M_t$ is a right-continuous martingale or positive submartingale,
and ${\mathbb E}[|M_{T}|^p]<\infty$ for some $p\geq1$, then
\begin{eqnarray}
{\mathbb E}\left[ \sup_{0\leq t\leq T}|M_t|^p \right] \leq \left(
\frac{p}{p-1}\right)^p {\mathbb E}\left[|M_{T}|^p\right],
\quad\quad (p>1)  , \label{eq:33.1}
\end{eqnarray}
and
\begin{eqnarray}
{\rm Prob}\left[ \sup_{0\leq t\leq T} |M_t| > \kappa \right] \leq
\frac{1}{\kappa^p} {\mathbb E}[|M_T|^p] , \quad\quad (p\geq1)
\label{eq:33.7}
\end{eqnarray}
for any constant $\kappa>0$.

In the present context, we are especially interested in the
inequality obtained in the case $p=2$, for which we have the
relation
\begin{eqnarray}
{\mathbb E}\left[\sup_{0\leq t\leq T}M_t^2\right] \leq 4 {\mathbb
E}\left[M_{T}^2\right] , \label{eq:33.2}
\end{eqnarray}
which is known as Doob's $L^2$-inequality. Now, setting
$M_t=H_t-H_0$ and using equation (\ref{eq:3.20}) we obtain
\begin{eqnarray}
{\mathbb E}\left[\sup_{0\leq t\leq T}(H_t-H_0)^2\right] \leq 4
(V_0-{ V}_T) . \label{eq:33.3}
\end{eqnarray}
In particular, taking the limit $T\rightarrow\infty$, it follows
from (\ref{eq:3.161}) that
\begin{eqnarray}
{\mathbb E}\left[\sup_{0\leq t\leq\infty}(H_t-H_0)^2\right] \leq 4
V_0 , \label{eq:33.4}
\end{eqnarray}
which shows that, on average, the energy stays within two standard
deviations of its original value.

This result is consistent with the intuition often arising in
physical arguments to the effect that when a system is in a state
of uncertain energy, then the energy fluctuates, with a typical
fluctuation roughly of the magnitude $\sim\Delta H$. There is no
quantum-mechanical principle which states that such fluctuations
actually occur, but one can see that in a martingale model there
may indeed be a natural basis for inferring the existence of
fluctuations of the required magnitude. We note that the bound
implied by the inequality (\ref{eq:33.4}) is independent of the
choice of $\sigma$, which shows that it is valid also for
relatively stable, long-lived states, i.e., those for which
$\sigma^2 V_0$ is small.

>From (\ref{eq:33.7}) we can determine an upper bound on the
probability that the magnitude of the energy fluctuation will
exceed any designated threshold during the reduction process.
Specifically, if we set $p=2$ and $\kappa=\lambda\sqrt{V_0}$, then
taking the limit $T\rightarrow\infty$ we obtain
\begin{eqnarray}
{\rm Prob}\left[ \sup_{0\leq t\leq\infty} (H_t-H_0)^2 > \lambda^2
V_0 \right] \leq \frac{1}{\lambda^2} . \label{eq:33.8}
\end{eqnarray}

A related bound for the variance process $V_t$ can be obtained by
use of Doob's maximal inequality for positive supermartingales
(see, e.g., \cite{yor}, p. 58). This relation states that, if
$X_t$ is a right-continuous positive supermartingale, then for any
constant $k\geq0$, we have
\begin{eqnarray}
{\rm Prob}\left[ \sup_{0\leq t\leq\infty} X_t > k \right] \leq
\frac{1}{k} {\mathbb E}[X_0] . \label{eq:33.5}
\end{eqnarray}
In the case of the variance process $V_t$, which as we have shown
is a positive supermartingale, if we set $k=\lambda^2 V_0$, then
(\ref{eq:33.5}) becomes
\begin{eqnarray}
{\rm Prob}\left[ \sup_{0\leq t\leq\infty}V_t>\lambda^2 V_0 \right]
\leq \frac{1}{\lambda^2} . \label{eq:33.6}
\end{eqnarray}
This relation shows that, during the reduction process, although
the energy variance can increase owing to random fluctuations,
there is a bound on the probability that the energy uncertainty
ever reaches $\lambda$ times the ensemble average of the initial
uncertainty for any given value of $\lambda$, and this bound is
given by $\lambda^{-2}$.

Let us return now to the asymptotic relation ${\mathbb
E}[H_\infty]=H_0$ and ask whether the terminal value $H_\infty$ of
the energy process actually exists as a random variable. To prove
that it does, we make use of the {\it martingale convergence
theorem}, which in a form sufficient for our purpose states that
if a continuous martingale $M_t$ satisfies ${\mathbb E}\left[
|M_t|^p \right]\leq k$ for some $p>1$ and $k<\infty$, and for all
$t\in[0,\infty)$, then there exists a random variable $M_\infty$
satisfying ${\mathbb E}\left[ |M_\infty|^p \right]\leq k$ and
$M_t={\mathbb E}_t[M_\infty]$, with the properties that
$\lim_{t\rightarrow\infty} M_t = M_\infty$ almost surely and that
$\lim_{t\rightarrow\infty} {\mathbb E}\left[ |M_t-M_\infty|^p
\right] = 0$.

In the present context, by setting $M_t=H_t-H_0$, we thus deduce
the existence of an asymptotic random variable $H_\infty$ with the
property that $H_t$ converges to $H_\infty$ almost surely, and
$H_t={\mathbb E}_t[H_\infty]$.

We conclude this section by generalising (\ref{eq:3.21}) to
demonstrate that the energy variance process $V_t$ defined by
(\ref{eq:3.6}) has the natural interpretation
\begin{eqnarray}
V_t = {\rm Var}_t\left[ H_\infty \right] .
\end{eqnarray}
That is to say, $V_t$ is given by the {\it conditional variance of
the terminal value of the energy, given information up to time
$t$}. To establish this relation we proceed as follows.

For any random variable $X$ on the probability space
$(\Omega,{\mathcal F},{\mathbb P})$ we define the conditional
variance ${\rm Var}[X|{\mathcal E}]$ with respect to the
$\sigma$-subfield ${\mathcal E}\subset{\mathcal F}$ by
\begin{eqnarray}
{\rm Var}[X|{\mathcal E}] = {\mathbb E}\left[(X-{\mathbb E}[X|
{\mathcal E}])^2|{\mathcal E}\right] .
\end{eqnarray}
It follows as an application of the law of total probability that
\begin{eqnarray}
{\rm Var}[X] = {\mathbb E}\left[ {\rm Var}[X|{\mathcal E}] \right]
+ {\rm Var}\left[{\mathbb E}[X|{\mathcal E}] \right] ,
\end{eqnarray}
the so-called {\sl conditional variance formula}. Thus for example
if ${\mathcal F}_t$ $(0\leq t<\infty)$ is a filtration of
$(\Omega,{\mathcal F},{\mathbb P})$ and we write ${\rm Var}_t[X] =
{\rm Var}[X|{\mathcal F}_t]$, then
\begin{eqnarray}
{\rm Var}_t[X] = {\mathbb E}_t\left[(X-{\mathbb E}_t[X])^2 \right]
\end{eqnarray}
and for the conditional variance formula we have
\begin{eqnarray}
{\rm Var}[X] = {\mathbb E}\left[ {\rm Var}_t[X] \right] + {\rm
Var}\left[{\mathbb E}_t[X] \right] .
\end{eqnarray}

In the problem at hand, we note that in the limit
$t\rightarrow\infty$ formula (\ref{eq:3.11}) for the variance
process $V_t$ takes the form
\begin{eqnarray}
V_0 + \sigma \int_0^\infty \beta_u {\rm d}W_u = \sigma^2
\int_0^\infty V_u^2 {\rm d}u .
\end{eqnarray}
Therefore, taking the conditional expectation of each side of this
relation and using formula (\ref{eq:3.75}) we deduce that
\begin{eqnarray}
V_0 + \sigma\int_0^t \beta_u {\rm d}W_u = \sigma^2 {\mathbb E}_t
\left[\int_0^\infty V_u^2 {\rm d}u \right] .
\end{eqnarray}
Substituting this relation into (\ref{eq:3.11}) then gives us
\begin{eqnarray}
V_t &=& \sigma^2{\mathbb E}_t \left[\int_0^\infty V_u^2 {\rm d}u
\right] -\sigma^2 \int_0^t V_u^2 {\rm d}u \nonumber \\
&=& \sigma^2{\mathbb E}_t \left[\int_t^\infty V_u^2 {\rm d}u
\right] \nonumber \\ &=& \sigma^2{\mathbb E}_t \left[ \left(
\int_t^\infty V_u {\rm d}W_u \right)^2\right] \\ &=&
{\mathbb E}_t \left[ (H_\infty-H_t)^2\right] \nonumber \\ &=& {\rm
Var}_t\left[ H_\infty\right], \nonumber
\end{eqnarray}
as desired. We note that in the next to last step here we have
used (\ref{eq:3.7}) together with the conditional Ito isometry
\begin{eqnarray}
{\mathbb E}_t \left[ \left( \int_t^T A_u {\rm d}W_u\right)^2
\right] = {\mathbb E}_t\left[ \int_t^T A_u^2 {\rm d}u\right],
\end{eqnarray}
valid for any adapted integrand $A_u$ satisfying ${\mathbb E}
\left[ \int_0^T A_u^2 {\rm d}u\right]<\infty$.

A positive supermartingale with the property that its expectation
goes to zero asymptotically is called a {\sl potential}
\cite{meyer}. The analysis above shows that the variance process
associated with quantum state reduction satisfies these conditions
and admits a {\sl Doob-Meyer decomposition} of the form
\begin{eqnarray}
V_t = {\mathbb E}_t\left[ Z_\infty\right] - Z_t ,
\end{eqnarray}
where
\begin{eqnarray}
Z_t=\sigma^2 \int_0^t V_u^2 {\rm d}u
\end{eqnarray}
is an increasing process.

\section{Reduction Probability}

The probability $\pi_n$ of reduction to a specific energy level
$E_n$, under the dynamics governed by the stochastic differential
equation (\ref{eq:3.1}), can be determined as follows. The method
we use is essentially that of \cite{grw,adler2}.

First we observe that for any operator ${\hat G}$ acting on
${\mathcal H}$, the process $G_t$ for the expectation value of
${\hat G}$ in the state $|\psi_t\rangle$ satisfies
\begin{eqnarray}
{\rm d} G_t &=& - {\rm i} \langle{\psi}_t|[{\hat G},{\hat
H}]|\psi_t \rangle {\rm d}t +\quat \sigma^2 \langle{\psi}_t|
\left( {\hat H}{\hat G}{\hat H} - \half\{{\hat H}^2,{\hat G}\}
\right)|\psi_t\rangle{\rm d}t \nonumber \\ & & +\half \sigma
\langle{\psi}_t| \{ ({\hat G}-G_t),({\hat H}-H_t)\} |\psi_t\rangle
{\rm d}W_t . \label{eq:4.1}
\end{eqnarray}
Here $[{\hat X},{\hat Y}]={\hat X}{\hat Y}-{\hat Y}{\hat X}$ and
$\{{\hat X},{\hat Y}\}={\hat X}{\hat Y}+{\hat Y}{\hat X}$ denote
the commutator and the anticommutator, respectively.

The drift term in (\ref{eq:4.1}) consists of two parts: the first
is the familiar Ehrenfest term involving the commutator with the
Hamiltonian; the second is a term of the Lindblad type $[{\hat
H},[{\hat G},{\hat H}]]$ arising as a consequence of the diffusive
dynamics of the state vector. The volatility term in
(\ref{eq:4.1}), i.e., the coefficient of ${\rm d}W_t$, is given by
the covariance of ${\hat G}$ and ${\hat H}$ in the state
$|\psi_t\rangle$. If ${\hat G}$ and ${\hat H}$ commute, then the
drift vanishes, and (\ref{eq:4.1}) reduces to
\begin{eqnarray}
{\rm d} G_t = \sigma \left( \langle{\psi}_t| {\hat G}{\hat H}
|\psi_t\rangle -G_tH_t \right) {\rm d}W_t , \label{eq:4.2}
\end{eqnarray}
from which it follows that the process $G_t$ is a martingale
\cite{adler2}. This is consistent with our earlier observation
that $H_t$ is itself a martingale, and that the process $G_t$
corresponding to any function of the form ${\hat G}=g({\hat H})$
is also a martingale.

Now let us consider the projection operator ${\hat P}_n$ for the
subspace ${\mathcal H}_n$ of ${\mathcal H}$ spanned by the energy
eigenstates with energy $E_n$. In the case of a nondegenerate
eigenvalue, we have ${\hat P}_n = |n\rangle \langle n|$. On the
other hand, if $E_n$ is a degenerate eigenvalue, then
\begin{eqnarray}
{\hat P}_n = \sum_{j=1}^{d_n} |n,j\rangle\langle n,j|,
\end{eqnarray}
as in (\ref{eq:2.7}), where $d_n$ is the dimension of the subspace
${\mathcal H}_n$ and $|n,j\rangle$ $(j=1,2,\cdots,d_n)$ is an
orthonormal basis for ${\mathcal H}_n$. Clearly ${\hat P}_n$
commutes with the Hamiltonian ${\hat H}$ for any value of $n$.
Furthermore, the relations ${\hat H}{\hat P}_n = {\hat P}_n{\hat
H} = E_n {\hat P}_n$ and ${\hat H} = \sum_n E_n {\hat P}_n$ are
equivalent on account of the resolution of identity
\begin{eqnarray}
\sum_{n=1}^{D} {\hat P}_n = 1 , \label{eq:4.7}
\end{eqnarray}
where $D$ is the number of distinct energy eigenvalues.

Now let us write
\begin{eqnarray}
P_{nt} = \frac{\langle{\psi}_t|{\hat P}_n|\psi_t\rangle}
{\langle{\psi}_t|\psi_t\rangle}  \label{eq:4.8}
\end{eqnarray}
for the expectation of the projection operator ${\hat P}_n$ in the
state $|\psi_t\rangle$. Because ${\hat P}_n$ commutes with the
Hamiltonian, we deduce that the process ${\hat P}_{nt}$ is a
martingale for each value of $n$. We note that $\sum_n P_{nt} = 1$
and $\sum_n E_n P_{nt} = H_t$. In particular, by setting
$G_t=P_{nt}$, one infers from (\ref{eq:4.2}) that
\begin{eqnarray}
{\rm d}P_{nt} = \sigma P_{nt} (E_n-H_t) {\rm d}W_t .
\label{eq:4.9}
\end{eqnarray}
This stochastic differential equation implies that $P_{nt}$ will
continue to fluctuate as long as $H_t\neq E_n$ and $P_{nt}\neq0$.
The solution of (\ref{eq:4.9}) is given by $P_{nt} = P_{n0}
M_{nt}$, where
\begin{eqnarray}
M_{nt} = \exp\left( \sigma \int_0^t (E_n-H_s) {\rm d} W_s - \half
\sigma^2 \int_0^t (E_n-H_s)^2 {\rm d}s \right) , \label{eq:4.10}
\end{eqnarray}
and $P_{n0}$ is the initial expectation value of the projection
operator ${\hat P}_n$. This follows from the fact that for any
bounded ${\mathcal F}_t$-adapted process $\sigma_t$ the solution
of the stochastic differential equation ${\rm d}X_t = \sigma_t X_t
{\rm d}W_t$ $(X_0>0)$ is
\begin{eqnarray}
X_t = X_0 \exp\left( \int_0^t \sigma_s {\rm d}W_s - \half \int_0^t
\sigma_s^2 {\rm d}s \right) , \label{eq:4.102}
\end{eqnarray}
which one can verify by an application of Ito's lemma. Because
$P_{nt}$ is a martingale, it follows that
\begin{eqnarray}
{\mathbb E}\left[P_{n\infty}\right] = P_{n0} . \label{eq:4.11}
\end{eqnarray}
Here ${\mathbb E}[P_{n\infty}]$ is the ensemble average of the
expectation value of the projection operator ${\hat P}_n$ at the
terminal energy eigenstate of the reduction process. Because
$P_{n\infty}$ takes the value one if the terminal energy has
eigenvalue $E_n$ and takes the value zero otherwise, it follows
that ${\mathbb E}[P_{n\infty}]$ is the {\it probability of
reaching a state with energy} $E_n$, i.e.,
\begin{eqnarray}
{\mathbb E}\left[P_{n\infty}\right] = \pi_n . \label{eq:4.12}
\end{eqnarray}

With these observations at hand, we are now in a position to
interpret the asymptotic martingale relation (\ref{eq:4.11}). If
$E_n$ is a nondegenerate eigenvalue, then $P_{n0}$ is the usual
expression for the Dirac transition probability from the initial
state $|\psi_0\rangle$ to the eigenstate $|n\rangle$, given by
\begin{eqnarray}
P_{n0} = \frac{\langle{\psi}_0|n\rangle\langle n|\psi_0\rangle}
{\langle{\psi}_0|\psi_0\rangle\langle n|n\rangle} .
\label{eq:4.13}
\end{eqnarray}
Thus we conclude, in the case of a nondegenerate Hamiltonian, that
the martingale model for quantum state reduction allows one to
{\it deduce} the correct transition probabilities.

In the case of a degenerate eigenstate, the probability $\pi_n$
can also be interpreted in terms of a Dirac transition
probability. In particular, whether or not the spectrum of the
Hamiltonian is degenerate, we can write
\begin{eqnarray}
P_{n0} &=& \frac{\langle{\psi}_0|{\hat P}_n|\psi_0\rangle}
{\langle{\psi}_0|\psi_0\rangle} \nonumber \\ &=&
\frac{(\langle{\psi}_0|{\hat P}_n|\psi_0\rangle)^2}
{\langle{\psi}_0|\psi_0\rangle\langle{\psi}_0|{\hat
P}_n|\psi_0\rangle} \nonumber \\ &=& \frac{\langle
P_n\psi_0|\psi_0\rangle \langle{\psi}_0|P_n\psi_0\rangle}
{\langle{\psi}_0|\psi_0\rangle \langle{\psi}_0|{\hat
P}_n|\psi_0\rangle} \label{eq:4.14} \\ &=& \frac{\langle P_n
\psi_0|\psi_0\rangle \langle{\psi}_0|P_n\psi_0\rangle}
{\langle{\psi}_0|\psi_0\rangle \langle{\psi}_0|{\hat
P}_n^2|\psi_0\rangle} \nonumber \\ &=& \frac{\langle P_n
\psi_0|\psi_0\rangle \langle{\psi}_0|P_n\psi_0\rangle}
{\langle{\psi}_0|\psi_0\rangle \langle P_n\psi_0|P_n\psi_0\rangle}
\nonumber ,
\end{eqnarray}
where the L\"uders state $|P_n\psi\rangle$ is defined by
\begin{eqnarray}
|P_n\psi_0\rangle \triangleq {\hat P}_n |\psi_0\rangle .
\label{eq:4.15}
\end{eqnarray}
Therefore, by virtue of (\ref{eq:4.11}) and (\ref{eq:4.14}), we
see that {\it the probability of obtaining the eigenvalue $E_n$ is
given by the Dirac transition probability from the given initial
state $|\psi_0\rangle$ to the L\"uders state $|P_n\psi_0\rangle$}.

The interesting point here is that, once again, while this is an
{\it assumption} in standard quantum theory, it arises as a {\it
deduction} in the martingale model for quantum state reduction.

In fact, we can demonstrate, in the case of a degenerate
eigenvalue, that the reduction necessarily results in the L\"uders
state if the corresponding eigenvalue is obtained. This can be
seen as follows.

For each value of $n=1,2,\cdots,D$, such that ${\hat P}_n|\psi_0
\rangle\neq0$ let us write
\begin{eqnarray}
|n,1\rangle = \frac{{\hat P}_n|\psi_0\rangle}
{\langle{\psi}_0|{\hat P}_n|\psi_0\rangle^{1/2}} \label{eq:4.16}
\end{eqnarray}
for a basis vector corresponding to the normalised L\"uders state
for that projection operator, and let $|n,j\rangle$, $j\neq1$, be
an associated basis for the states orthogonal to $|n,1\rangle$
that lie in the subspace ${\mathcal H}_n$ spanned by eigenstates
of energy $E_n$. The operator
\begin{eqnarray}
{\hat \Pi}_n &\triangleq& \sum_{j=2}^{d_n}|n,j\rangle\langle n,j|
\nonumber \\ &=& {\hat P}_n - |n,1\rangle\langle n,1|
\label{eq:4.17}
\end{eqnarray}
thus projects onto the subspace of ${\mathcal H}_n$ consisting of
vectors {\it orthogonal} to the L\"uders state for that value of
$n$. Evidently, we have
\begin{eqnarray}
{\hat\Pi}_n|\psi_0\rangle = 0 \label{eq:4.18}
\end{eqnarray}
which follows from (\ref{eq:4.16}) and the fact that
${\hat\Pi}_n{\hat P}_n={\hat\Pi}_n$. Since the projection operator
${\hat\Pi}_n$ commutes with the Hamiltonian, the process
\begin{eqnarray}
\Pi_{nt} = \frac{\langle{\psi}_t|{\hat\Pi}_n|\psi_t\rangle}
{\langle{\psi}_t|\psi_t\rangle} \label{eq:4.19}
\end{eqnarray}
is a martingale, the initial value of which is $\Pi_{n0}=0$ on
account of the relation (\ref{eq:4.18}). Therefore, by virtue of
the martingale relation ${\mathbb E} \left[ \Pi_{n\infty} \right]
=\Pi_{n0}$, we deduce that ${\mathbb E}[\Pi_{n\infty}]=0$. Now,
$\Pi_{n\infty}$ is a nonnegative random variable. Therefore, if
${\mathbb E}[\Pi_{n\infty}]=0$ then $\Pi_{n\infty}=0$ almost
surely. It follows that the terminal state must be orthogonal to
the subspace of ${\mathcal H}_n$ spanned by states with energy
$E_n$ that are orthogonal to the L\"uders state. As a consequence,
we see that {\it if reduction occurs to a state of energy $E_n$,
then that state must be the L\"uders state corresponding to that
eigenvalue}.

In fact, we deduce a stronger result: namely, that {\it the
stochastic motion of the state vector, during the course of the
reduction process, is necessarily confined to the $D$-dimensional
subspace of ${\mathcal H}$ spanned by the L\"uders states ${\hat
P}_n|\psi_0\rangle$, for $n=1,2,\cdots,D$, where $D$ is the number
of distinct energy eigenvalues and ${\hat P}_n$ is the projection
operator onto the subspace ${\mathcal H}_n$ of ${\mathcal H}$
spanned by eigenstates with eigenvalue $E_n$}.

The proof of this theorem follows from the fact that, for each
$n$, the process $\Pi_{nt}$ is a martingale, and because
$\Pi_{n0}=0$ we have ${\mathbb E}[\Pi_{nt}]=0$ for all $t\geq0$
and thus $\Pi_{nt}=0$ for all $t\geq0$. Therefore the state vector
$|\psi_t\rangle$ always lies in the space spanned by the vectors
${\hat P}_n|\psi_0\rangle$ for $n=1,2,\cdots,D$.

A similar analysis is valid in the more general situation for
which the dynamics of $|\psi_t\rangle$ are given by a stochastic
differential equation of the form
\begin{eqnarray}
{\rm d}|\psi_t\rangle &=& -{\rm i}{\hat H}|\psi_t\rangle {\rm d}t
- \octa \sum_{\alpha=1}^{r} \sigma_\alpha^2 ({\hat
F}_\alpha-F_{\alpha t})^2 |\psi_t\rangle {\rm d}t \nonumber \\ & &
+ \half \sum_{\alpha=1}^{r} \sigma_\alpha ({\hat
F}_\alpha-F_{\alpha t}){\rm d}W_t^\alpha .
\end{eqnarray}
Here ${\hat F}_\alpha$ $(\alpha=1,2,\cdots,r)$ represents a
commuting family of observables, each of which also commutes with
the Hamiltonian ${\hat H}$, the $\sigma_\alpha$ are associated
coupling constants, and we write $F_{\alpha t}$ for the
expectation of ${\hat F}_{\alpha}$ in the state $|\psi_t\rangle$.
In this case $W_t^{\alpha}$ denotes a standard $r$-dimensional
Brownian motion, and the reduction proceeds to a common eigenstate
of operators ${\hat F}_\alpha$ $(\alpha=1,2,\cdots,r)$. Most of
the results of this paper are applicable {\it mutatis mutandis} to
this more general class of reduction process, though in what
follows we shall, for simplicity, continue to confine the detailed
discussion to the case of the energy-based reduction
(\ref{eq:3.1}).

\section{The Case of an Initially Mixed State}

Thus far we have considered the role of the L\"uders postulate as
it applies to an initially {\sl pure} state $|\psi_0\rangle$, and
we have demonstrated that the postulate follows directly as a
consequence of the martingale model for quantum state reduction.
The L\"uders postulate is, however, applicable in a more general
context as well: namely, when the initial state is specified as a
mixture with density matrix ${\hat\rho}_0$. In that case, when an
observable $F$ is measured, the associated state reduction is
given by the L\"uders rule
\begin{eqnarray}
{\hat\rho}_0 \mapsto \frac{{\hat P}_n{\hat\rho}_0{\hat P}_n} {{\rm
Tr}{\hat P}_n{\hat\rho}_0 } , \label{eq:5.1}
\end{eqnarray}
for the density matrix, if the measurement result is the
eigenvalue $f_n$, and this occurs with probability
\begin{eqnarray}
\pi_n = {\rm Tr} {\hat P}_n {\hat\rho}_0 . \label{eq:5.2}
\end{eqnarray}
Here, as before, ${\hat P}_n$ denotes the projection operator onto
the subspace ${\mathcal H}_n$ of ${\mathcal H}$ spanned by
eigenstates with the eigenvalue $f_n$, which may or may not be
degenerate.

The interpretation of an expression involving density matrices,
such as (\ref{eq:5.1}), is best understood in terms of ensemble
averages. Thus (\ref{eq:5.1}) means that if initially
${\hat\rho}_0$ can be used to compute the expectation of any
observable $G$, not necessarily compatible with $F$, then after
$F$ is measured, and if the result $f_n$ is observed, the density
matrix ${\hat P}_n {\hat\rho}_0 {\hat P}_n/{\rm Tr}{\hat P}_n
{\hat\rho}_0$ can be used to compute the expectation of $G$ in a
subsequent measurement.

Additionally, given the initial density matrix ${\hat\rho}_0$, if
$F$ is measured but no note is taken of the result, then the
ensemble average for a subsequent measurement of the observable
$G$ is ${\rm Tr}{\hat\rho}_\infty{\hat G}$, where
\begin{eqnarray}
{\hat\rho}_\infty = \sum_n {\hat P}_n {\hat\rho}_0 {\hat P}_n .
\label{eq:5.3}
\end{eqnarray}
It should be borne in mind that these expressions, while generally
regarded as part of the standard apparatus of quantum theory, are
not derivable from any of the more basic assumptions of quantum
mechanics, and have to be regarded as constituting an additional
postulate. See, e.g., \cite{isham} for an illuminating brief
account of the status of the projection postulate in quantum
mechanics, and its relation to state reduction. It is interesting
to note that von Neumann, in his original splendid work on the
subject \cite{neumann}, apparently failed to offer a satisfactory
expression for the density matrix in the case of the measurement
of an observable with a degenerate spectrum, a deficiency only
later rectified by L\"uders and others \cite{luders} (cf.
\cite{pauli}, section 9, and the remark attributed to A.~S.
Wightman on p. 550 of ref. \cite{zurek}).

The general L\"uders rule (\ref{eq:5.1}) has the important
property that, in the measurement of an observable with a
degenerate spectrum, if the initial state is not pure, then the
final state need not be pure, if the result of the measurement is
one of the degenerate eigenvalues.

Now let us see if we can gain a clearer understanding of the
general L\"uders formulae (\ref{eq:5.1}), (\ref{eq:5.2}) and
(\ref{eq:5.3}) by consideration of the martingale model for
quantum state reduction. In the theory of stochastic differential
equations, it is acceptable that the initial value of the random
process should itself be a random variable; thus it is merely a
special case when $|\psi_0\rangle$ in (\ref{eq:3.1}) is known. The
deterministic case corresponds to the situation where the initial
density matrix is pure, i.e., of rank one. In the general case,
where $|\psi_0\rangle$ is random, i.e., given by a mixture, the
corresponding initial density matrix ${\hat\rho}_0$ is the
ensemble average
\begin{eqnarray}
{\hat\rho}_0 = {\mathbb E}\left[ |{\bf\Psi}_0\rangle\langle
{\bf\Psi}_0|\right] , \label{eq:5.4}
\end{eqnarray}
where $|{\bf\Psi}_0\rangle$ is a random initial state vector,
which we assume to be normalised. Then for the final density
matrix we have
\begin{eqnarray}
{\hat\rho}_\infty = \sum_n {\mathbb E}\left[ \pi_n({\bf\Psi}_0)
\frac{{\hat P}_n|{\bf\Psi}_0\rangle\langle{\bf\Psi}_0|{\hat P}_n}
{\langle{\bf\Psi}_0|{\hat P}_n|{\bf\Psi}_0\rangle} \right]
\label{eq:5.5}
\end{eqnarray}
as a consequence of the reduction
\begin{eqnarray}
|{\bf\Psi}_0\rangle \mapsto \frac{{\hat P}_n|{\bf\Psi}_0\rangle}
{\langle{\bf\Psi}_0|{\hat P}_n|{\bf\Psi}_0\rangle^{1/2}} ,
\label{eq:5.6}
\end{eqnarray}
where $\pi_n({\bf\Psi}_0)$ is the conditional probability that the
eigenvalue $E_n$ is obtained, given the random initial state
$|{\bf\Psi}_0\rangle$. However, this conditional probability is
given by
\begin{eqnarray}
\pi_n({\bf\Psi}_0) = \langle{\bf\Psi}_0|{\hat P}_n
|{\bf\Psi}_0\rangle , \label{eq:5.7}
\end{eqnarray}
i.e., the Dirac transition probability to the random L\"uders
state determined by the random initial state
$|{\bf\Psi}_0\rangle$, in accordance with (\ref{eq:4.14}). As a
consequence we see that (\ref{eq:5.5}) simplifies to give
\begin{eqnarray}
{\hat\rho}_\infty &=& \sum_n {\mathbb E}\left[ {\hat P}_n|
{\bf\Psi}_0\rangle\langle{\bf\Psi}_0|{\hat P}_n \right] \nonumber
\\ &=& \sum_n {\hat P}_n{\mathbb E}\left[ |{\bf\Psi}_0\rangle
\langle{\bf\Psi}_0| \right]{\hat P}_n \label{eq:5.8} \\
&=& \sum_n {\hat P}_n {\hat\rho}_0 {\hat P}_n , \nonumber
\end{eqnarray}
and thus we obtain (\ref{eq:5.3}). Additionally we have
\begin{eqnarray}
{\hat\rho}_{\infty} = \sum_n \pi_n {\hat\rho}_{n\infty} ,
\label{eq:5.9}
\end{eqnarray}
where
\begin{eqnarray}
{\hat\rho}_{n\infty} = \frac{{\hat P}_n{\hat\rho}_0{\hat P}_n}
{{\rm Tr}{\hat P}_n{\hat\rho}_0}  \label{eq:5.10}
\end{eqnarray}
is the reduced or `conditional' density matrix, given that the
observer has knowledge of the result $H=E_n$, and
\begin{eqnarray}
\pi_n &=& {\mathbb E}\left[ \pi_n({\bf\Psi}_0) \right] \nonumber
\\ &=& {\mathbb E}\left[ \langle{\bf\Psi}_0|{\hat P}_n|
{\bf\Psi}_0\rangle \right] \nonumber \\ &=& {\mathbb E} \left[
{\rm Tr}{\hat P}_n |{\bf\Psi}_0\rangle \langle{\bf\Psi}_0| \right]
\label{eq:5.11} \\ &=& {\rm Tr}{\hat P}_n {\mathbb E} \left[
|{\bf\Psi}_0\rangle\langle{\bf\Psi}_0| \right] \nonumber \\ &=&
{\rm Tr}{\hat P}_n {\hat\rho}_0 \nonumber
\end{eqnarray}
is the probability of this result.

\section{Dynamics of the Density Matrix}

To gain further insight into the case where the initial density
matrix is not pure, we can make a computation of the dynamics for
${\hat\rho}_t$. This can be achieved by examination of the
Lindblad equation associated with the stochastic differential
equation (\ref{eq:3.1}), which in this case turns out to be
solvable.

If we start with equation (\ref{eq:4.1}) for the dynamics of the
expectation value of an arbitrary operator ${\hat G}$ in the state
$|\psi_t\rangle$, and take the ensemble average
\begin{eqnarray}
{\mathbb E}\left[ G_t\right] = {\rm Tr}{\hat G}{\hat\rho}_t ,
\label{eq:6.1}
\end{eqnarray}
where ${\hat\rho}_t={\mathbb E}[|\psi_t\rangle\langle\psi_t|]$, we
find that
\begin{eqnarray}
{\rm d}{\mathbb E}[G_t] &=& {\rm Tr}{\hat G}{\rm d}{\hat\rho}_t
\nonumber \\ &=& -{\rm i}{\rm Tr}{\hat\rho}_t [{\hat G},{\hat
H}]{\rm d}t +\quat\sigma^2 {\rm Tr} {\hat\rho}_t \left({\hat
H}{\hat G}{\hat H}-\half{\hat H}^2{\hat G} - \half
{\hat G}{\hat H}^2\right){\rm d}t \label{eq:6.2} \\
&=& -{\rm i}{\rm Tr}{\hat G}[{\hat H},{\hat\rho}_t]{\rm d}t +
\quat\sigma^2 {\rm Tr}{\hat G} \left({\hat H}{\hat\rho}_t{\hat H}
-\half{\hat\rho}_t{\hat H}^2-\half {\hat H}^2{\hat\rho}_t
\right){\rm d}t , \nonumber
\end{eqnarray}
where in the second equality we make use of the cyclic property of
the trace operator. This relation has to hold for any observable
$G$, from which it follows that
\begin{eqnarray}
\partial_t{\hat\rho}_t = -{\rm i}[{\hat H},{\hat\rho}_t] + \quat\sigma^2 
\left(
{\hat H}{\hat\rho}_t{\hat H} - \half {\hat H}^2{\hat\rho}_t -\half
{\hat\rho}_t{\hat H}^2 \right) , \label{eq:6.3}
\end{eqnarray}
where $\partial_t=\partial/\partial t$. This is the general
equation of the Lindblad type \cite{gorini,lindblad} associated
with the stochastic differential equation (\ref{eq:3.1}), for
which $\half\sigma{\hat H}$ is the corresponding Lindblad
operator.

Now we consider the problem of solving the Lindblad equation
(\ref{eq:6.3}) subject to an arbitrary specification of the
initial density matrix ${\hat\rho}_0$. For convenience we switch
to a Heisenberg representation in which the density matrix is
defined by the operator
\begin{eqnarray}
{\hat r}_t \triangleq {\rm e}^{{\rm i}{\hat H}t}{\hat\rho}_t {\rm
e}^{-{\rm i}{\hat H}t} . \label{eq:6.4}
\end{eqnarray}
This has the effect of removing the purely unitary part of the
evolution. For the dynamics of ${\hat r}_t$ we have
\begin{eqnarray}
\partial_t{\hat r}_t = {\rm e}^{{\rm i}{\hat H}t}(\partial_t{\hat\rho}_t)
{\rm e}^{-{\rm i}{\hat H}t} + {\rm i}[{\hat H},{\hat r}_t] ,
\label{eq:6.5}
\end{eqnarray}
and therefore
\begin{eqnarray}
\partial_t{\hat r}_t = \quat \sigma^2 \left(
{\hat H}{\hat r}_t{\hat H} - \half {\hat H}^2
{\hat r}_t - \half {\hat r}_t {\hat H}^2 \right) . \label{eq:6.6}
\end{eqnarray}
Let us write ${\hat P}_n$ for the projection operator onto the
subspace ${\mathcal H}_n$. Then, because ${\hat P}_n {\hat H} =
{\hat H} {\hat P}_n = E_n {\hat P}_n$, if we multiply each side of
equation (\ref{eq:6.6}) by ${\hat P}_n$ on both the right and the
left we obtain
\begin{eqnarray}
\partial_t({\hat P}_n{\hat r}_t{\hat P}_n) = 0 , \label{eq:6.7}
\end{eqnarray}
from which it follows that ${\hat P}_n{\hat r}_t{\hat P}_n$ is a
constant of the motion. In particular, we have ${\hat P}_n{\hat
r}_0{\hat P}_n={\hat P}_n{\hat r}_\infty{\hat P}_n$, and thus
${\hat P}_n{\hat\rho}_0{\hat P}_n={\hat P}_n{\hat\rho}_\infty
{\hat P}_n$, and therefore
\begin{eqnarray}
\sum_n {\hat P}_n{\hat\rho}_0{\hat P}_n = \sum_n {\hat P}_n {\hat
\rho}_\infty {\hat P}_n . \label{eq:6.8}
\end{eqnarray}
Because the terminal state is necessarily a mixture of energy
eigenstates we have
\begin{eqnarray}
\sum_n{\hat P}_n{\hat\rho}_\infty{\hat P}_n = {\hat\rho}_\infty,
\label{eq:6.9}
\end{eqnarray}
from which by use of (\ref{eq:6.8}) we immediately infer the
general form of the L\"uders reduction postulate (\ref{eq:5.3}).

To proceed further we define the operator matrix valued process
${\hat{\mathcal R}}_{nm}(t)$ by
\begin{eqnarray}
{\hat{\mathcal R}}_{nm}(t) \triangleq {\hat P}_n{\hat r}_t{\hat
P}_m . \label{eq:6.10}
\end{eqnarray}
For each of the values of $n$ and $m$, ${\hat {\mathcal
R}}_{nm}(t)$ is a time-dependent Hermitian operator. Here,
$n,m=1,2,\cdots,D$, where $D$ is the number of distinct energy
levels. From equation (\ref{eq:6.6}) for the dynamics of ${\hat
r}_t$ we deduce, by use of the relation ${\hat P}_n{\hat
H}=E_n{\hat P}_n$, that
\begin{eqnarray}
\partial_t{\hat{\mathcal R}}_{nm}(t)=-\octa\sigma^2(E_n-E_m)^2{\hat
{\mathcal R}}_{nm}(t) . \label{eq:6.11}
\end{eqnarray}
The general solution of the ordinary differential equation
(\ref{eq:6.11}) is given by
\begin{eqnarray}
{\hat{\mathcal R}}_{nm}(t) &=& {\hat{\mathcal R}}_{nm}(0)
\exp\left( -\octa \sigma^2(E_n-E_m)^2t \right) \nonumber \\ &=&
{\hat P}_n{\hat r}_0{\hat P}_m \exp\left( -\octa
\sigma^2(E_n-E_m)^2t \right) . \label{eq:6.12}
\end{eqnarray}

On the other hand, by use of the resolution of the identity
(\ref{eq:4.7}) it follows from (\ref{eq:6.10}) that
\begin{eqnarray}
{\hat r}_t &=& \sum_{n,m} {\hat{\mathcal R}}_{nm}(t) \nonumber
\\ &=& \sum_{n,m} {\hat P}_n {\hat r}_0 {\hat
P}_m \exp\left( -\octa\sigma^2(E_n-E_m)^2t\right) \label{eq:6.13} \\
&=& \sum_{n\neq m} {\hat P}_n {\hat r}_0 {\hat P}_m \exp
\left(-\octa\sigma^2(E_n-E_m)^2t\right)  + \sum_n {\hat P}_n {\hat
r}_0 {\hat P}_n . \nonumber
\end{eqnarray}
Therefore, by inverting the transformation (\ref{eq:6.4}), we
obtain the solution of the Lindblad equation in the original
Schr\"odinger picture as:
\begin{eqnarray}
{\hat\rho}_t = \sum_{n\neq m} {\hat P}_n {\hat\rho}_0 {\hat P}_m
{\rm e}^{-{\rm i}(E_n-E_m)t-\octa\sigma^2(E_n-E_m)^2t} + \sum_n
{\hat P}_n {\hat\rho}_0 {\hat P}_n . \label{eq:6.14}
\end{eqnarray}
We recover the initial state ${\hat\rho}_0$ by setting $t=0$ in
the right hand side of (\ref{eq:6.14}). The off-diagonal terms are
damped away exponentially at the rate $\octa\sigma^2V_{nm}$ as
$t\rightarrow\infty$, where $V_{nm}=(E_n-E_m)^2$ is the square of
the spread between relevant energy levels, and we are left with
the L\"uders state (\ref{eq:5.3}) for the terminal density matrix
${\hat\rho}_\infty$.

It is interesting to observe that the L\"uders state, obtained by
the limit as $t\rightarrow\infty$ of the density matrix associated
with the reduction process (\ref{eq:3.1}), coincides with the
asymptotic time average of the density matrix in the case of a
purely unitary evolution governed by the von Neumann equation
\begin{eqnarray}
\frac{\partial{\hat\rho}_t}{\partial t} = -{\rm i} [{\hat H},
{\hat\rho}_t] , \label{eq:6.15}
\end{eqnarray}
for which the solution is ${\hat\rho}_t= {\rm e}^{-{\rm i}{\hat
H}t}{\hat\rho}_0{\rm e}^{{\rm i}{\hat H}t}$. More specifically, if
we write
\begin{eqnarray}
\langle{\hat\rho}\rangle_T \triangleq \frac{1}{T} \int_0^T {\hat
\rho}_t {\rm d}t \label{eq:6.16}
\end{eqnarray}
for the time average of ${\hat\rho}_t$ up to time $T$, then we
find that
\begin{eqnarray}
\lim_{T\rightarrow\infty}\langle{\hat\rho}\rangle_T = \sum_{n}
{\hat P}_n {\hat\rho}_0 {\hat P}_n , \label{eq:6.17}
\end{eqnarray}
where ${\hat P}_n$ is the projection operator onto the subspace
${\mathcal H}_n\subset{\mathcal H}$ spanned by the states of
energy $E_n$. This result can be verified directly by use of the
resolution of the identity (\ref{eq:4.7}). The calculation is as
follows:
\begin{eqnarray}
\langle{\hat\rho}\rangle_T &=& \frac{1}{T} \sum_{m,n} \int_0^T
{\hat P}_n {\rm e}^{-{\rm i}{\hat H}t} {\hat\rho}_0 {\rm e}^{{\rm
i}{\hat H}t} {\hat P}_m {\rm d}t \nonumber \\ &=& \frac{1}{T}
\sum_{m,n}{\hat P}_n {\hat\rho}_0 {\hat P}_m
\int_0^T {\rm e}^{-{\rm i}(E_n-E_m)t} {\rm d}t \label{eq:6.18} \\
&=& \sum_n {\hat P}_n {\hat\rho}_0 {\hat P}_n + \frac{1}{T}
\sum_{m\neq n} {\hat P}_n {\hat\rho}_0 {\hat P}_m \left(
\frac{\sin(\omega_{nm}T)}{\omega_{nm}} + {\rm i}
\frac{\cos(\omega_{nm}T)-1}{\omega_{nm}} \right) , \nonumber
\end{eqnarray}
where $\omega_{nm}=E_n-E_m$. Therefore, in the limit
$T\rightarrow\infty$ the off-diagonal terms drop out, and we
recover (\ref{eq:6.17}). For a closely related result see
\cite{parthasarathy}.

\section{Change of Measure}

We return now to the stochastic differential equation
(\ref{eq:3.1}) governing quantum state reduction with a view to
gaining further insights into the nature of the resulting
dynamics. We shall demonstrate in this section how a `change of
measure' technique can be used to solve (\ref{eq:3.1}) and thus,
in effect, to construct an explicit unravelling of the Lindblad
equation (\ref{eq:6.3}). The general problem of formulating an
appropriate unravelling of the Lindblad equation in a given
physical context is a matter of considerable interest in a number
of areas of modern physics
\cite{carmichael,diosi3,brun2,adler3,gardiner,wiseman}.

We begin with the following remark. Let ${\hat\mu}_t$ and
${\hat\sigma}_t$ be bounded ${\mathcal F}_t$-adapted
operator-valued processes on $(\Omega,{\mathcal F},{\mathbb P})$
with the property that for all $s,t\in[0, \infty)$ the random
matrices ${\hat\mu}_s$, ${\hat\mu}_t$, ${\hat\sigma}_s$ and
${\hat\sigma}_t$ mutually commute. Then the stochastic
differential equation
\begin{eqnarray}
{\rm d}|\psi_t\rangle ={\hat\mu}_t|\psi_t\rangle {\rm d}t +
{\hat\sigma}_t |\psi_t\rangle {\rm d}W_t \label{eq:10.1}
\end{eqnarray}
has the unique solution
\begin{eqnarray}
|\psi_t\rangle = \exp\left( \int_0^t \left( {\hat\mu}_s-\half
{\hat\sigma}^2_s\right) {\rm d}s + \int_0^t{\hat\sigma}_s {\rm
d}W_s \right) |\psi_0\rangle . \label{eq:10.2}
\end{eqnarray}
Here we allow for the possibility that the initial state
$|\psi_0\rangle$ may be random. A straightforward application of
Ito's lemma shows that (\ref{eq:10.2}) leads back to
(\ref{eq:10.1}). In the case of the reduction process
(\ref{eq:3.1}), which is evidently of the form (\ref{eq:10.1}), we
can write
\begin{eqnarray}
{\hat\mu}_t = -{\rm i}{\hat H}-\octa\sigma^2 \left({\hat H} - H_t
\right)^2 , \quad \quad {\hat\sigma}_t = \half \sigma \left({\hat
H} - H_t \right). \label{eq:10.4}
\end{eqnarray}
It follows therefore that
\begin{eqnarray}
|\psi_t\rangle &=& \exp\left( -{\rm i}{\hat H}t - \quat\sigma^2
\int_0^t \left({\hat H}-H_s\right)^2{\rm d}s \right. \nonumber \\
& & \quad\quad\quad\quad\quad\quad\quad\quad \left. + \half \sigma
\int_0^t\left({\hat H}-H_s\right) {\rm d}W_s \right)
|\psi_0\rangle . \label{eq:10.5}
\end{eqnarray}
This is still an implicit solution for $|\psi_t\rangle$, because
$H_s=\langle\psi_s|{\hat H}|\psi_s\rangle$. Nevertheless, as a
consequence of (\ref{eq:10.5}) we see that the evolution of the
state vector according to (\ref{eq:3.1}) can be expressed in the
simple form
\begin{eqnarray}
|\psi_t\rangle = {\hat U}_t {\hat R}_t |\psi_0\rangle ,
\label{eq:10.6}
\end{eqnarray}
where the operator-valued process ${\hat U}_t$ is defined by
\begin{eqnarray}
{\hat U}_t \triangleq \exp\left( -{\rm i}{\hat H}t\right) ,
\label{eq:10.7}
\end{eqnarray}
and the operator-valued process ${\hat R}_t$ is defined by
\begin{eqnarray}
{\hat R}_t \triangleq \exp\left( \half \sigma \int_0^t\left({\hat
H}-H_s\right) {\rm d}W_s - \quat\sigma^2 \int_0^t \left({\hat
H}-H_s\right)^2{\rm d}s \right). \label{eq:10.8}
\end{eqnarray}
We note that ${\hat U}_t$ is unitary and that ${\hat U}_t$ and
${\hat R}_t$ commute. The square of ${\hat R}_t$, which we denote
by ${\hat M}_t$, is an operator-valued martingale. The fact that
${\hat M}_t$ satisfies the martingale condition ${\mathbb
E}_s[{\hat M}_t]={\hat M}_s$ is evident from the expression
\begin{eqnarray}
{\hat M}_t = \exp\left( \sigma \int_0^t\left({\hat H}-H_s\right)
{\rm d}W_s - \half \sigma^2 \int_0^t \left({\hat H} -H_s \right)^2
{\rm d}s \right). \label{eq:10.9}
\end{eqnarray}
In particular, if ${\hat P}_n$ is the projection operator onto the
subspace ${\mathcal H}_n\subset{\mathcal H}$ spanned by the states
of energy $E_n$, then we find that
\begin{eqnarray}
{\hat M}_t = \sum_n {\hat P}_n M_{nt},  \label{eq:10.91}
\end{eqnarray}
where $M_{nt}$ is given by (\ref{eq:4.10}). We note that for each
value of $n$ the process $M_{nt}$ is an exponential martingale.

Now suppose that ${\hat G}$ is an observable that commutes with
the Hamiltonian ${\hat H}$. Then for its expectation in the state
$|\psi_t\rangle$ we have
\begin{eqnarray}
G_t &=& \langle\psi_t|{\hat G}|\psi_t\rangle \nonumber \\
&=& \langle\psi_0|{\hat R}_t{\hat U}_t^{\dagger}{\hat G} {\hat
U}_t {\hat R}_t |\psi_0\rangle \label{eq:10.11} \\
&=& \langle\psi_0|{\hat G} {\hat M}_t |\psi_0\rangle , \nonumber
\end{eqnarray}
and therefore
\begin{eqnarray}
{\mathbb E}_s\left[G_t\right] &=& {\mathbb E}_s\left[ \langle
\psi_0|{\hat G} {\hat M}_t |\psi_0\rangle\right] \nonumber \\
&=& \langle\psi_0|{\hat G} {\mathbb E}_s [ {\hat M}_t ]
|\psi_0\rangle \nonumber \\ &=& \langle\psi_0|{\hat G}{\hat M}_s
|\psi_0\rangle \label{eq:10.12} \\ &=& \langle\psi_0| {\hat R}_s
{\hat G}{\hat R}_s|\psi_0\rangle \nonumber \\ &=& \langle\psi_0|
{\hat R}_s {\hat U}_s^\dagger {\hat G}{\hat U}_s {\hat
R}_s|\psi_0\rangle \nonumber \\ &=& \langle\psi_s| {\hat G}
|\psi_s\rangle, \nonumber
\end{eqnarray}
which shows that $G_t$ is a martingale. In this way we are able to
verify directly that the dynamical law (\ref{eq:3.1}) implies that
the expectation value of any observable that commutes with the
Hamiltonian is a weakly conserved quantity.

To proceed further we note that it is a straightforward algebraic
exercise to verify that ${\hat M}_t$ can be expressed as the
following quotient:
\begin{eqnarray}
{\hat M}_t = \frac{\exp\left( \sigma \int_0^t {\hat H} ({\rm
d}W_s+\sigma H_s{\rm d}s) - \frac{1}{2} \sigma^2 \int_0^t {\hat
H}^2 {\rm d}s \right)} {\exp\left( \sigma \int_0^t H_s ({\rm d}
W_s+\sigma H_s{\rm d}s) - \frac{1}{2} \sigma^2 \int_0^t H_s^2 {\rm
d}s \right)} . \label{eq:10.13}
\end{eqnarray}
In particular, let us define the `modified' Brownian motion
process $W^*_t$ by
\begin{eqnarray}
W_t^* \triangleq W_t + \sigma \int_0^t H_s {\rm d}s,
\label{eq:10.14}
\end{eqnarray}
so ${\rm d}W_t^*={\rm d}W_t+H_t{\rm d}t$. Then, because ${\hat H}$
is constant, we can write ${\hat M}_t$ in the simple form
\begin{eqnarray}
{\hat M}_t = \frac{1}{\Lambda_t^*} \exp\left( \sigma {\hat H}
W_t^* - \half \sigma^2 {\hat H}^2 t \right), \label{eq:10.15}
\end{eqnarray}
where
\begin{eqnarray}
\Lambda_t^* \triangleq \exp\left( \sigma \int_0^t H_s {\rm d}W_s^*
- \half \sigma^2 \int_0^t H_s^2 {\rm d}s \right) .
\label{eq:10.16}
\end{eqnarray}
The significance of the processes $W_t^*$ and $\Lambda_t^*$ will
become apparent shortly.

We have already verified that (\ref{eq:3.1}) preserves the norm of
$|\psi_0\rangle$. If we assume that $\langle
\psi_0|\psi_0\rangle=1$, then it follows from (\ref{eq:10.6}) that
$\langle\psi_0|{\hat M}_t|\psi_0\rangle=1$ for all $t$. Thus we
deduce from (\ref{eq:10.13}) and (\ref{eq:10.14}) that
\begin{eqnarray}
\Lambda_t^* = \langle\psi_0| \exp\left( \sigma {\hat H} W_t^* -
\half \sigma^2 {\hat H}^2 t\right) |\psi_0\rangle .
\label{eq:10.17}
\end{eqnarray}
As a consequence we can write
\begin{eqnarray}
{\hat M}_t = \frac{\exp\left( \sigma {\hat H}W_t^* - \half
\sigma^2 {\hat H}^2 t\right)}{\langle\psi_0| \exp\left( \sigma
{\hat H}W_t^* - \half \sigma^2 {\hat H}^2 t\right) |\psi_0
\rangle} , \label{eq:10.18}
\end{eqnarray}
which has the important effect of localising the dependence of
${\hat M}_t$ on $H_t$ in the modified Brownian motion $W_t^*$. The
process $H_t$ in turn is given by (\ref{eq:3.2}), from which it
follows that $H_t = \langle\psi_0|{\hat H}{\hat M}_t|\psi_0
\rangle$. Therefore, by use of (\ref{eq:10.18}) we have
\begin{eqnarray}
H_t = \frac{\langle{\psi}_0| {\hat H} \exp\left( \sigma {\hat H}
W_t^* - \frac{1}{2} \sigma^2 {\hat H}^2 t\right) |\psi_0\rangle}
{\langle{\psi}_0| \exp\left( \sigma{\hat H} W_t^* - \frac{1}{2}
\sigma^2 {\hat H}^2 t\right) |\psi_0\rangle} ,\label{eq:10.19}
\end{eqnarray}
which shows that $H_t$ can be expressed as a function of $W_t^*$
and $t$. This is given explicitly by
\begin{eqnarray}
H_t = \frac{\sum_n \pi_n E_n \exp\left( \sigma E_n W_t^* - \half
\sigma^2 E_n^2 t\right)}{\sum_n \pi_n \exp\left( \sigma E_n W_t^*
-\half\sigma^2 E_n^2 t\right)} , \label{eq:10.20}
\end{eqnarray}
where as usual $\pi_n$ denotes the probability that the eigenvalue
attained is $E_n$, given the initial state $|\psi_0\rangle$. We
also note that
\begin{eqnarray}
\Lambda_t^* = \sum_n \pi_n \exp\left( \sigma E_n W_t^* - \half
\sigma^2 E_n^2 t \right) , \label{eq:10.21}
\end{eqnarray}
and that
\begin{eqnarray}
{\hat M}_t = \frac{\sum_n {\hat P}_n \exp\left( \sigma E_n W_t^* -
\half \sigma^2 E_n^2 t\right)}{\sum_n \pi_n \exp\left( \sigma E_n
W_t^* -\half\sigma^2 E_n^2 t\right)} . \label{eq:10.22}
\end{eqnarray}

Now we proceed to examine the processes $W_t^*$ and $\Lambda_t^*$
more closely. This is the point at which we introduce the highly
useful concept of a {\sl change of probability measure}. We shall
see in what follows that there exists a change of measure
${\mathbb P}\rightarrow {\mathbb Q}$ such that for any given
interval of time $[0,T]$ the process $W_t^*$ for $t\in[0,T]$ is a
Brownian motion with respect to the probability space
$(\Omega,{\mathcal F}_T,{\mathbb Q})$ and the filtration
${\mathcal F}_t$ $(0\leq t\leq T)$. The implication of this is
that with respect to the measure ${\mathbb Q}$ the basic processes
$H_t$, $\Lambda_t^*$, and ${\hat M}_t$ can be expressed in terms
of ratios of sums of geometric Brownian motions, thus offering a
significant element of analytic tractability.

We begin with a few mathematical preliminaries concerning the
change of measure technique. Given the probability space
$(\Omega,{\mathcal F},{\mathbb P})$, we recall that $1_{A}$
denotes the indicator function of the event $A\in{\mathcal F}$.
Thus for each $\omega\in\Omega$ we have $1_{A}(\omega)=1$ if
$\omega\in A$ and $1_{A}(\omega)=0$ if $\omega\notin A$. It
follows that
\begin{eqnarray}
{\rm Prob}^{\mathbb P}[A] = {\mathbb E}^{\mathbb P} \left[ 1_{A}
\right] , \label{eq:10.23}
\end{eqnarray}
where ${\rm Prob}^{\mathbb P}$ and ${\mathbb E}^{\mathbb P}$
denote probability and expectation with respect to the measure
${\mathbb P}$.

Now let $\Lambda$ be a positive random variable on the probability
space $(\Omega,{\mathcal F},{\mathbb P})$. Then we can define a
new probability measure ${\mathbb Q}$ on the underlying measurable
space $(\Omega,{\mathcal F})$ by the formula
\begin{eqnarray}
{\rm Prob}^{\mathbb Q}[A] = \frac{{\mathbb E}^{\mathbb P}[\Lambda
1_{A}]}{{\mathbb E}^{\mathbb P}[\Lambda]} . \label{eq:10.24}
\end{eqnarray}
Because $\Lambda$ is positive, this relation is invertible and we
have
\begin{eqnarray}
{\rm Prob}^{\mathbb P}[A] = \frac{{\mathbb E}^{\mathbb Q}[
\Lambda^* 1_{A}]}{{\mathbb E}^{\mathbb Q}[\Lambda^*]} ,
\label{eq:10.25}
\end{eqnarray}
where $\Lambda^*=1/\Lambda$. The two probability measures
${\mathbb P}$ and ${\mathbb Q}$ in this case are said to be {\sl
equivalent} in the sense that they agree on null sets, i.e., for
all $A\in{\mathcal F}$ we have ${\rm Prob}^{\mathbb P}[A]=0$ if
and only if ${\rm Prob}^{\mathbb Q}[A]=0$.

In the case of a filtered probability space some important
additional structure arises in this connection. Suppose the
process $\Lambda_t$ is a positive martingale on $(\Omega,{\mathcal
F},{\mathbb P})$ with respect to the filtration ${\mathcal F}_t$,
satisfying $\Lambda_0=1$. For any fixed value of $t$ the random
variable $\Lambda_t$ can be used to define a measure ${\mathbb
Q}_t$ on $(\Omega,{\mathcal F}_t)$ according to the procedure
outlined in the previous paragraph. It follows then from
(\ref{eq:10.24}) by virtue of the martingale property of
$\Lambda_t$ that
\begin{eqnarray}
{\rm Prob}^{{\mathbb Q}_t} [A] = {\mathbb E}^{\mathbb P}\left[
\Lambda_t 1_{A} \right] \label{eq:10.26}
\end{eqnarray}
for all $A\in{\mathcal F}_t$. We note that if $s\leq t$ then ${\rm
Prob}^{{\mathbb Q}_s}[A] ={\rm Prob}^{{\mathbb Q}_t}[A]$ for all
$A\in{\mathcal F}_s$. This is because
\begin{eqnarray}
{\rm Prob}^{{\mathbb Q}_t}[A] &=& {\mathbb E}^{\mathbb P}\left[
\Lambda_t 1_{A}\right] \nonumber \\ &=& {\mathbb E}^{\mathbb P}
\left[{\mathbb E}^{\mathbb P}[\Lambda_t 1_{A}|{\mathcal F}_s]
\right] \nonumber \\ &=& {\mathbb E}^{\mathbb P} \left[{\mathbb
E}^{\mathbb P}[\Lambda_t |{\mathcal F}_s]1_{A} \right]
\label{eq:10.27} \\ &=& {\mathbb E}^{\mathbb P} \left[ \Lambda_s
1_{A}\right] \nonumber \\ &=& {\rm Prob}^{{\mathbb Q}_s}[A] .
\nonumber
\end{eqnarray}
Therefore, for any finite interval of time $[0,T]$ the measure
thus obtained on $(\Omega,{\mathcal F}_T)$ is independent of the
specific choice of $T$. Thus we can drop the suffix on ${\mathbb
Q}$ and speak of the change of measure ${\mathbb P} \rightarrow
{\mathbb Q}$ induced by the given density martingale $\Lambda_t$.

The key result making use of this apparatus that we require in
what follows is the theorem of Girsanov (see, e.g.,
\cite{karatzas}). Let $[0,T]$ be a fixed interval of time, and
$W_t$ a Brownian motion on $(\Omega,{\mathcal F}_T)$ with respect
to the filtration ${\mathcal F}_t$ $(0\leq t\leq T)$ and the
measure ${\mathbb P}$. Suppose that the process $\lambda_t$ is
${\mathcal F}_t$-adapted and that
\begin{eqnarray}
\Lambda_t = \exp\left( -\int_0^t \lambda_s {\rm d}W_s - \half
\int_0^t \lambda_s^2 {\rm d}s \right) . \label{eq:10.29}
\end{eqnarray}
is a martingale. Then Girsanov's theorem states that the modified
process
\begin{eqnarray}
W_t^* \triangleq W_t + \int_0^t \lambda_s {\rm d}s
\label{eq:10.28}
\end{eqnarray}
is a Brownian motion with respect to the equivalent measure
${\mathbb Q}$ induced by the density martingale $\Lambda_t$. For
$\Lambda_t$ to be a martingale it suffices that $\lambda_s$ should
satisfy the Novikov condition
\begin{eqnarray}
{\mathbb E}^{\mathbb P}\left[ \exp\left( \half\int_0^T \lambda_s^2
{\rm d}s \right)\right] < \infty . \label{eq:10.30}
\end{eqnarray}
In particular, if $\lambda_t$ is bounded, then $\Lambda_t$ is a
martingale.

If $\Lambda_t$ is a ${\mathbb P}$-martingale then the associated
process $\Lambda_t^*=1/\Lambda_t$ given by
\begin{eqnarray}
\Lambda_t^* = \exp\left( +\int_0^t \lambda_s {\rm d}W_s^* - \half
\int_0^t \lambda_s^2 {\rm d}s \right) , \label{eq:10.31}
\end{eqnarray}
is a ${\mathbb Q}$-martingale, and induces the inverse change of
measure ${\mathbb Q}\rightarrow{\mathbb P}$. In particular, for
any ${\mathcal F}_t$-measurable random variable $X_t$ we have the
following formulae for the calculation of expectations:
\begin{eqnarray}
{\mathbb E}_s^{\mathbb P}[X_t] = \frac{1}{\Lambda_s^*} {\mathbb
E}_s^{\mathbb Q}\left[ \Lambda_t^* X_t\right] , \label{eq:10.311}
\end{eqnarray}
and its reversal
\begin{eqnarray}
{\mathbb E}_s^{\mathbb Q}[X_t] = \frac{1}{\Lambda_s} {\mathbb
E}_s^{\mathbb P}\left[ \Lambda_t X_t\right] . \label{eq:10.312}
\end{eqnarray}

Returning to the matter at hand, we note that for quantum state
reduction the process $\lambda_t$ is given by $\sigma H_t$, and
the corresponding change of measure density martingale is given by
\begin{eqnarray}
\Lambda_t = \exp\left( -\sigma \int_0^t H_s {\rm d}W_s - \half
\sigma^2\int_0^t H_s^2 {\rm d}s \right)  \label{eq:10.32}
\end{eqnarray}
for the transformation ${\mathbb P}\rightarrow{\mathbb Q}$. The
process $W_t^*$ as defined by (\ref{eq:10.14}) is a ${\mathbb
Q}$-Brownian motion. The associated inverse transformation
${\mathbb Q}\rightarrow{\mathbb P}$ is induced by the process
$\Lambda_t^*$ defined in (\ref{eq:10.16}).

Now we are in a position to give a complete characterisation of
the solution of the dynamical equation (\ref{eq:3.1}) for the
state reduction problem valid over any finite time interval
$[0,T]$. The recipe is as follows.

We start with the measure ${\mathbb Q}$ for which $W_t^*$ is a
Brownian motion. Given $W_t^*$ we then construct the process $H_t$
by use of formula (\ref{eq:10.20}), and the process $\Lambda_t^*$
by use of formula (\ref{eq:10.21}), and the process ${\hat M}_t$
by use of formula (\ref{eq:10.22}). Thus we see that the wave
function $|\psi_t\rangle$ along with all the related processes
${\hat R}_t$, ${\hat M}_t$, $\Lambda_t$, and $H_t$ can be
explicitly constructed as functions of $W_t^*$ and $t$. The
physical measure ${\mathbb P}$ constructed by use of $\Lambda_t^*$
is then used for the calculation of ensemble averages. In
particular, letting ${\mathbb E}$ denote the expectation with
respect to the physical measure ${\mathbb P}$, it follows from
(\ref{eq:10.22}) that
\begin{eqnarray}
{\mathbb E}\left[ X_t\right] = {\mathbb E}^{\mathbb Q}\left[
\Lambda_t^* X_t\right] \label{eq:10.165}
\end{eqnarray}
for any ${\mathcal F}_t$-measurable random variable $X_t$.

For example, suppose ${\hat G}$ is an observable that does not
necessarily commute with the Hamiltonian ${\hat H}$, and we wish
to calculate the ensemble average of the expectation value
$\langle \psi_t|{\hat G}|\psi_t\rangle$. Then by use of
(\ref{eq:10.165}) we have
\begin{eqnarray}
{\mathbb E}\left[\langle\psi_t|{\hat G}|\psi_t\rangle\right] &=&
{\mathbb E}^{\mathbb Q}\left[\Lambda_t^*\langle\psi_t|{\hat G}
|\psi_t\rangle\right] \nonumber \\ &=& {\mathbb E}^{\mathbb Q}
\left[\Lambda_t^*\langle\psi_0|{\hat U}_t^\dagger{\hat R}_t{\hat
G}{\hat R}_t {\hat U}_t|\psi_0\rangle\right] \label{eq:10.33} \\
&=& {\mathbb E}^{\mathbb Q} \left[ \langle\psi_0|{\rm e}^{{\rm
i}{\hat H}t+\frac{1}{2}\sigma {\hat H} W_t^* -
\frac{1}{4}\sigma^2{\hat H}^2 t}{\hat G}{\rm e}^{-{\rm i}{\hat H}t
+\frac{1}{2}\sigma {\hat H} W_t^* - \frac{1}{4} \sigma^2{\hat H}^2
t}|\psi_0\rangle\right] \nonumber \\ &=& {\mathbb E}^{\mathbb Q}
\left[\sum_{m,n} G_{mn} {\rm e}^{{\rm i}
(E_m-E_n)t+\frac{1}{2}\sigma(E_m+E_n)W_t^* -\frac{1}{4} (E_m^2 +
E_n^2)\sigma^2 t}\right] , \nonumber
\end{eqnarray}
where the matrix elements $G_{mn}$ are given by
\begin{eqnarray}
G_{mn} = \langle\psi_0|{\hat P}_m{\hat G}{\hat P}_n|\psi_0\rangle.
\label{eq:10.34}
\end{eqnarray}
Since $W_t^*$ is normally distributed with mean zero and variance
$t$ with respect to the ${\mathbb Q}$-measure, the expectation in
(\ref{eq:10.33}) can be readily computed. By use of the simple
relation
\begin{eqnarray}
{\mathbb E}^{\mathbb Q}\left[ {\rm e}^{\alpha W_t^*}\right] = {\rm
e}^{\frac{1}{2}\alpha^2 t} , \label{eq:10.35}
\end{eqnarray}
which holds for any constant $\alpha$, we see that
\begin{eqnarray}
{\mathbb E}^{\mathbb Q} \left[ {\rm e}^{ \frac{1}{2}
\sigma(E_m+E_n)W_t^* - \frac{1}{4}(E_m^2 + E_n^2)\sigma^2
t}\right] = {\rm e} ^{-\frac{1}{8}\sigma^2(E_m-E_n)^2 t} .
\label{eq:10.36}
\end{eqnarray}
As a consequence we deduce that
\begin{eqnarray}
{\mathbb E}\left[\langle\psi_t|{\hat G}|\psi_t\rangle\right] =
\sum_{m,n} {\bar G}_{mn} {\rm e}^{{\rm i}(E_m-E_n)t -\frac{1}{8}
\sigma^2(E_m-E_n)^2 t} , \label{eq:10.37}
\end{eqnarray}
where
\begin{eqnarray}
{\bar G}_{mn} &=& {\mathbb E}\left[ \langle\psi_0| {\hat P}_m
{\hat G}{\hat P}_n|\psi_0\rangle\right] \nonumber \\ &=& {\rm Tr}
{\hat\rho}_0({\hat P}_m {\hat G}{\hat P}_n) \\ &=& {\rm Tr} {\hat
G}({\hat P}_n {\hat\rho}_0{\hat P}_m) \nonumber
\end{eqnarray}
and ${\hat\rho}_0$ is the density matrix corresponding to the
random initial state. This result is consistent with our earlier
expression (\ref{eq:6.14}) for the solution of the Lindblad
equation, and illustrates the fact that the change of measure
technique is indeed highly effective as a calculational tool for
quantum state reduction models.

\vspace{0.5cm}
\begin{footnotesize}
\noindent SLA and TAB acknowledge support by DOE Grant No.
DE-FG02-90ER40542. DCB acknowledges support from The Royal
Society. LPH acknowledges the Institute for Advanced Study for
hospitality while part of this work was carried out. We are
grateful to E.~J.~Brody, L.~P.~Horwitz, B.~K.~Meister, and
K.~P.~Tod for stimulating discussions.

\noindent Electronic mail: ${}^1$adler@ias.edu
${}^2$dorje@ic.ac.uk ${}^3$tbrun@ias.edu
${}^4$lane.hughston@kcl.ac.uk
\end{footnotesize}


\vspace{0.5cm}

\begin{thebibliography}{999}

\bibitem{neumann} von Neumann,~J. {\em Mathematische Grundlagen
der Quantenmechanik} (Springer, Berlin 1932); translation into
English by Beyer,~R.~T., {\it Mathematical Foundations of Quantum
Mechanics} (Princeton University Press, Princeton 1971).

\bibitem{luders} L\"{u}ders,~G. Annalen der Physik {\bf 8}, 322
(1951).

\bibitem{gisin} Gisin,~N., Phys. Rev. Lett. {\bf 52}, 1657 (1984);
Helv. Phys. Acta {\bf 62}, 363 (1989).

\bibitem{diosi} Diosi,~L., J. Phys. A {\bf 21}, 2885 (1988); Phys.
Lett. A {\bf 129}, 419 (1988); Phys. Lett. A {\bf 132}, 233
(1988).

\bibitem{grw} Ghirardi,~G.C., Pearle,~P. and Rimini,~A.,
Phys. Rev. A {\bf 42}, 78 (1990).

\bibitem{percival} Percival,~I., Proc. R. Soc. London A {\bf 447}, 189
(1994).

\bibitem{hughston} Hughston,~L.~P., Proc. R. Soc. London A
{\bf 452}, 953 (1996).

\bibitem{adler2} Adler,~S.~L. and Horwitz,~L.~P., J. Math. Phys.
{\bf 41}, 2485 (2000).

\bibitem{brody} Brody,~D.~C. and Hughston,~L.~P., Preprint
(quant-ph/0011125).

\bibitem{pearle0} Pearle,~P., in {\it Open Systems and Measurement
in Relativistic Quantum Theory}, H.-P.~Breuer and F.~Petruccione,
eds. (Springer, Berlin 2000).

\bibitem{ghirardi} Ghirardi,~G.~C., in {\em Quantum Reflections},
J. Ellis and D. Amati, eds. (Cambridge University Press, Cambridge
2000).

\bibitem{hughston3} Hughston,~L.~P., Jozsa,~R. and Wooters,~W.~K.,
Phys. Lett. A {\bf 183}, 14 (1993).

\bibitem{adler} Adler,~S.~L. and Brun,~T.~A., J. Phys. A {\bf 34},
4797 (2001).

\bibitem{percival3} Percival,~I.~C., J. Phys. A{\bf 27}, 1003
(1994).

\bibitem{schack} Schack,~R., Brun,~T.~A. and Percival,~I.~C., J.
Phys. A{\bf 28}, 5401 (1995).

\bibitem{ikeda} Ikeda,~N. and Watanabe,~S. {\em Stochastic
Differential Equations and Diffusion Processes} (North-Holland,
Amsterdam 1981).

\bibitem{yor} Revuz,~D. and Yor,~M. {\em Continuous Martingales
and Brownian Motion}, 3rd ed., Corrected 2nd print (Springer,
Berlin 2001).

\bibitem{meyer} Meyer,~P.~A. {\em Probability and Potentials}
(Blaisdell Publishing Company, Waltham, Massachusetts 1966).

\bibitem{isham} Isham,~C.~J., {\em Lectures on Quantum Theory}
(Imperial College Press, London 1995).

\bibitem{pauli} Pauli,~W., ``Die allgemeinen Prinzipien der
Wellenmechanik,'' in {\em Handbuch der Physik}, H. Geiger and K.
Scheel, eds., {\bf 24}, pp. 83-272 (Springer-Verlag, Berlin,
1933).

\bibitem{zurek} Wheeler,~J.~A. and Zurek,~W.~H. (eds) {\em Quantum
Theory and Measurement} (Princeton University Press, Princeton
1983).

\bibitem{gorini} Gorini,~V., Kossakowski,~A. and
Sudarshan,~E.~C.~G., J. Math. Phys. {\bf 17}, 821 (1976).

\bibitem{lindblad} Lindblad,~G., Commun. Math. Phys. {\bf 48}, 119
(1976).

\bibitem{parthasarathy} Parthasarathy,~K.~R. {\em An Introduction
to Quantum Stochastic Calculus} (Birkh\"auser, Basel 1992).

\bibitem{carmichael} Carmichael,~H.~J. {\em An Open Systems
Approach to Quantum Optics} (Springer, Berlin 1993).

\bibitem{diosi3} Diosi,~L., Gisin,~N., Halliwell,~J. and
Percival,~I.~C., Phys. Rev. Lett. {\bf 24}, ... (1995).

\bibitem{brun2} Brun,~T.~A., Phys. Rev. Lett. {\bf 78}, 1833
(1997).

\bibitem{adler3} Adler,~S.~L., Phys. Lett. A{\bf 265}, 58 (2000).

\bibitem{gardiner} Gardiner,~C.~W. and Zoller,~P., {\em Quantum
Noise}, 2nd ed. (Springer, Berlin 2000).

\bibitem{wiseman} Wiseman,~H.~M. and Diosi,~L., J. Chem. Phys.
{\bf 268}, 91 (2001).

\bibitem{karatzas} Karatzas,~I. and Shreve,~S.~E. {\em Brownian
Motion and Stochastic Calculus}, 2nd ed. (Springer, Berlin 1991).



\end{thebibliography}
\end{document}